\documentclass[twocolumn]{aastex631} 

\usepackage{amsmath}	
\usepackage{amssymb}	
\usepackage{wasysym}
\usepackage{multirow}
\usepackage[flushleft]{threeparttable}
\usepackage{booktabs}
\usepackage{hyperref}
\usepackage{tablefootnote}

\newcolumntype{Y}{>{\centering\arraybackslash}X}
\usepackage{array}  
\turnoffediting

\usepackage{color}

\newcommand{\system}{BD+05\,4868}

\newcommand{\figr}[1]{Figure~\ref{fig:#1}}
\newcommand{\tabr}[1]{Table~\ref{table:#1}}

\shorttitle{Bright Star, Fading World}
\shortauthors{Hon et al.}

\graphicspath{{./}{figures/}}

\begin{document}
\newcommand{\PSUAA}{Department of Astronomy \& Astrophysics, 525 Davey Laboratory, The Pennsylvania State University, University Park, PA, 16802, USA}
\newcommand{\PSUCEHW}{Center for Exoplanets \& Habitable Worlds, 525 Davey Laboratory, The Pennsylvania State University, University Park, PA 16802, USA}
\newcommand{\PSETI}{Penn State Extraterrestrial Intelligence Center, 525 Davey Laboratory, The Pennsylvania State University, University Park, PA 16802, USA}
\newcommand{\UA}{Steward Observatory, The University of Arizona, 933 N.\ Cherry Ave, Tucson, AZ 85721, USA}
\newcommand{\Caltech}{Department of Astronomy, California Institute of Technology, Pasadena, CA 91125, USA}
\newcommand{\JHU}{Department of Physics \& Astronomy, Bloomberg Center, Johns Hopkins University, Baltimore, MD 21218, USA}
\newcommand{\Macquarie}{School of Mathematical and Physical Sciences, Macquarie University, Balaclava Road, North Ryde, NSW 2109, Australia}
\newcommand{\CUBoulder}{Department of Physics, 390 UCB, University of Colorado, Boulder, CO 80309, USA}
\newcommand{\JPL}{Jet Propulsion Laboratory, California Institute of Technology, 4800 Oak Grove Drive, Pasadena, CA 91109, USA}
\newcommand{\MITEAPS}{Department of Earth, Atmospheric, and Planetary Sciences, Massachusetts Institute of Technology, Cambridge, MA 02139, USA}
\newcommand{\MITKavli}{Kavli Institute for Astrophysics and Space Research, Massachusetts Institute of Technology, Cambridge, MA 02139, USA}
\newcommand{\MITAero}{Department of Aeronautics and Astronautics, Massachusetts Institute of Technology, 77 Massachusetts Avenue, Cambridge, MA 02139, USA}
\newcommand{\UCI}{Department of Physics \& Astronomy, The University of California, Irvine, Irvine, CA 92697, USA}
\newcommand{\Carnegie}{Earth and Planets Laboratory, Carnegie Institution for Science, 5241 Broad Branch Road, NW, Washington, DC 20015, USA}
\newcommand{\PSUICS}{Institute for Computational and Data Sciences, The Pennsylvania State University, University Park, PA 16802, USA}
\newcommand{\PSUCASt}{Center for Astrostatistics, 525 Davey Laboratory, The Pennsylvania State University, University Park, PA 16802, USA}
\newcommand{\Princeton}{Department of Astrophysical Sciences, Princeton University, 4 Ivy Lane, Princeton, NJ 08540, USA}
\newcommand{\IAS}{Institute for Advance Study, 1 Einstein Drive, Princeton, NJ 08540, USA}
\newcommand{\Tsinghua}{Department of Astronomy, Tsinghua University, Beijing 100084, China}
\newcommand{\FlatironCCA}{Center for Computational Astrophysics, Flatiron Institute, 162 Fifth Avenue, New York, NY 10010, USA}
\newcommand{\ETH}{ETH Zurich, Institute for Particle Physics \& Astrophysics, Zurich, Switzerland}
\newcommand{\UCO}{UC Observatories, University of California, Santa Cruz, CA 95064, USA}
\newcommand{\SantaCruz}{University of California, Santa Cruz}
\newcommand{\WMKO}{W.\ M.\ Keck Observatory, 65-1120 Mamalahoa Hwy, Kamuela, HI 96743, USA}
\newcommand{\SSL}{Space Sciences Laboratory, University of California, Berkeley, CA 94720, USA}
\newcommand{\UH}{Institute for Astronomy, University of Hawai‘i, 2680 Woodlawn Drive, Honolulu, HI 96822, USA}
\newcommand{\UCB}{Department of Astronomy, 501 Campbell Hall, University of California, Berkeley, CA 94720, USA}
\newcommand{\UCLA}{Department of Physics \& Astronomy, University of California Los Angeles, Los Angeles, CA 90095, USA}
\newcommand{\nexsci}{NASA Exoplanet Science Institute/Caltech-IPAC, California Institute of Technology, Pasadena, CA
91125, USA}
\newcommand{\COO}{Caltech Optical Observatories, California Institute of Technology, Pasadena, CA 91125, USA}
\newcommand{\Sydney}{Sydney Institute for Astronomy (SIfA), School of Physics, University of Sydney, NSW 2006, Australia}
\newcommand{\Kansas}{Department of Physics and Astronomy, University of Kansas, Lawrence, KS, USA}
\newcommand{\Warwick}{Physics Department, University of Warwick, Coventry CV4 7AL, United Kingdom}
\newcommand{\Yale}{Department of Astronomy, Yale University, New Haven, CT 06520, USA}
\newcommand{\ICL}{Astrophysics Group, Department of Physics, Imperial College London, Prince Consort Rd, London SW7 2AZ, UK}
\newcommand{\Schmidt}{Astrophysics \& Space Institute, Schmidt Sciences, New York, NY 10011, USA}
\newcommand{\Amsterdam}{Anton Pannekoek Institute for Astronomy, University of Amsterdam, Science Park 904, 1098 XH Amsterdam, The Netherlands}
\newcommand{\ND}{Department of Physics and Astronomy, University of Notre Dame, Notre Dame, IN 46556, USA}
\newcommand{\Geneva}{Observatoire Astronomique de l'Université de Genève, Chemin Pegasi 51, 1290 Versoix, Switzerland}
\newcommand{\WDRC}{Center for Solar-Stellar Connections, White Dwarf Research Corporation,  9020 Brumm Trail, Golden, CO 80403, USA}
\newcommand{\CEA}{Universit\'e Paris-Saclay, Universit\'e Paris Cit\'e, CEA, CNRS, AIM, F-91191, Gif-sur-Yvette, France}
\newcommand{\SAC}{Stellar Astrophysics Centre, Aarhus University, Ny Munkegade 120, DK-8000 Aarhus C, Denmark}
\newcommand{\PortoIAC}{Instituto de Astrof\'{i}sica e Ci\^{e}ncias do Espaço, Universidade do Porto, Rua das Estrelas, 4150-762 Porto, Portugal}
\newcommand{\Porto}{Departamento de F\'{i}sica e Astronomia, Faculdade de Ci\^{e}ncias da Universidade do Porto, Rua do Campo Alegre, s/n, 4169-007 Porto,
Portugal}
\newcommand{\UNSW}{School of Physics, University of New South Wales, NSW 2052, Australia}
\newcommand{\ASTROTHREED}{ARC Centre of Excellence for All Sky Astrophysics in Three Dimensions (ASTRO-3D)}
\newcommand{\CFA}{Center for Astrophysics $|$ Harvard \& Smithsonian, 60 Garden Street, Cambridge, MA 02138, USA}
\newcommand{\Liege}{Astrobiology Research Unit, Universit\'e de Li\`ege, 19C All\'ee du 6 Ao\^ut, 4000 Li\`ege, Belgium}
\newcommand{\TenerifeIAC}{Instituto de Astrof\'isica de Canarias (IAC), Calle V\'ia L\'actea s/n, 38200, La Laguna, Tenerife, Spain}
\newcommand{\Komaba}{Komaba Institute for Science, The University of Tokyo, 3-8-1 Komaba, Meguro, Tokyo 153-8902, Japan}
\newcommand{\AstroBioJapan}{Astrobiology Center, 2-21-1 Osawa, Mitaka, Tokyo 181-8588, Japan}
\newcommand{\Lowell}{Lowell Observatory, 1400 W Mars Hill Road, Flagstaff, AZ, 86001, USA}
\title{A Disintegrating Rocky Planet with Prominent Comet-like Tails Around a Bright Star}

\author[0000-0003-2400-6960]{Marc Hon}
\affiliation{\MITKavli}

\author[0000-0003-3182-5569]{Saul Rappaport}
\affiliation{\MITKavli}

\author[0000-0002-1836-3120]{Avi Shporer}
\affiliation{\MITKavli}

\author[0000-0001-7246-5438]{Andrew Vanderburg}
\affiliation{\MITKavli}

\author[0000-0001-6588-9574]{Karen A. Collins}
\affiliation{\CFA}
\author[0000-0001-8621-6731]{Cristilyn N.\ Watkins}
\affiliation{\CFA}

\author[0000-0001-8227-1020]{Richard P. Schwarz}
\affiliation{\CFA}

\author[0000-0003-1464-9276]{Khalid Barkaoui}
\affiliation{\Liege}
\affiliation{\MITEAPS}
\affiliation{\TenerifeIAC}

\author[0000-0001-7961-3907]{Samuel W. Yee}
\affil{\CFA}
\affil{\Princeton}

\author[0000-0002-4265-047X]{Joshua N. Winn}
\affil{\Princeton}

\author[0000-0001-7047-8681]{Alex S. Polanski}
\altaffiliation{Percival Lowell Fellow}
\affil{\Lowell}
\affil{\Kansas}

\author[0000-0002-0388-8004]{Emily A. Gilbert}
\affil{\JPL}

\author[0000-0002-5741-3047]{David R. Ciardi}
\affil{\nexsci}

\author[0000-0002-4371-3460]{Jeroen Audenaert}
\affil{\MITKavli}

\author[0000-0003-0241-2757]{William Fong}
\affil{\MITKavli}

\author[0009-0005-3397-2046]{Jack Haviland}
\affil{\MITKavli}

\author[0000-0002-2135-9018]{Katharine Hesse}
\affil{\MITKavli}

\author[0000-0002-5788-9280]{Daniel Muthukrishna}
\affil{\MITKavli}

\author[0009-0005-2122-0680]{Glen Petitpas}
\affil{\MITKavli}

\author[0000-0002-4853-8075]{Ellie Hadjiyska Schmelzer}
\affil{\MITKavli}

\author[0000-0001-8511-2981]{Norio Narita}
\affiliation{\Komaba}
\affiliation{\AstroBioJapan}
\affiliation{\TenerifeIAC}

\author[0000-0002-4909-5763]{Akihiko Fukui}
\affiliation{\Komaba}
\affiliation{\TenerifeIAC}

\author[0000-0002-6892-6948]{Sara Seager}
\affiliation{\MITKavli}
\affiliation{\MITEAPS}
\affiliation{\MITAero}

\author[0000-0003-2058-6662]{George R. Ricker}
\affiliation{\MITKavli}


\correspondingauthor{Marc Hon}
\email{mtyhon@mit.edu}

\begin{abstract}
We report the discovery of \system\,Ab, a transiting exoplanet orbiting a bright ($V=10.16$) K-dwarf (TIC 466376085) with a period of 1.27 days. Observations from NASA's Transiting Exoplanet Survey Satellite (TESS) reveal variable transit depths and asymmetric transit profiles that are characteristic of comet-like tails formed by dusty effluents emanating from a disintegrating planet. Unique to \system\,Ab is the presence of prominent dust tails in both the trailing and leading directions that contribute to the extinction of starlight from the host star. By fitting the observed transit profile and analytically modeling the drift of dust grains within both dust tails, we infer large grain sizes ($\sim1-10\,\mu$m) and a mass loss rate of $10\,M_{\rm \oplus}$\,Gyr$^{-1}$, suggestive of a lunar-mass object with a disintegration timescale of only several Myr. The host star is probably older than the Sun and is accompanied by an M-dwarf companion at a projected physical separation of 130 AU. The brightness of the host star, combined with the planet's relatively deep transits ($0.8-2.0\%$), presents \system\,Ab as a prime target for compositional studies of rocky exoplanets and investigations into the nature of catastrophically evaporating planets. 
\end{abstract}


\section{Introduction} \label{sec:intro}

\begin{figure}
    \centering
	\includegraphics[width=1.\columnwidth]{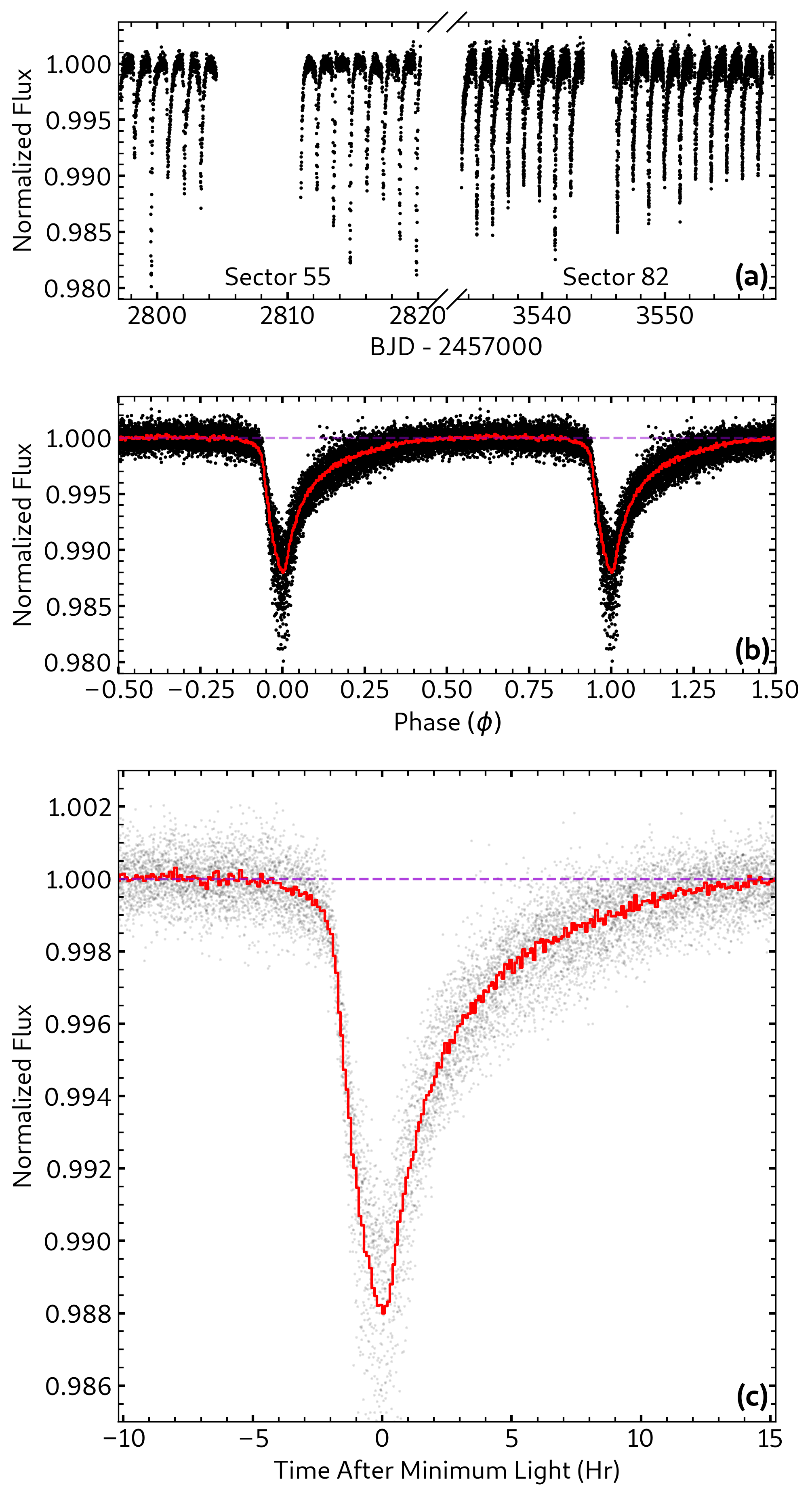}
    \caption{TESS light curve of \system\,A. (a) Normalized and detrended light curves from Sectors 55 and 82. (b) Phase-folded light curve, repeated for two periods. The light curve is folded at a period of 1.271869 days ($\sim30.5\,$hrs) and binned over 6-minute intervals (red), showing the average transit profile across 29 transits. (c) Close-up of the transit in the phase-folded light curve.}
    \label{fig:fig1}
\end{figure}

Small, rocky planets orbiting their host stars too closely can vaporize. Subject to tidal locking and extreme stellar irradiation, the day-side surfaces of such planets are hot enough to melt and develop thin atmospheres of evaporated rocky material (e.g., \citealt{Leger_2011, Kite_2016, Zilinkas_2022, Curry_2024}). The surface gravity of a hot, close-in planet with a mass comparable to that of Mercury, or smaller, is too weak to retain such an atmosphere. In such cases, the atmospheric vapor tends to escape into space through a thermally driven wind \citep{Rappaport_2012}, often referred to as a Parker-type wind (e.g., \citealt{Lamers_1999, Perez-Becker_2013}). Gas pressure accelerates the wind away from the planet, where the gas expands and cools. These conditions result in the condensation of dust grains that eventually decouple from the wind and drift away from the planet \citep{Perez-Becker_2013, Booth_2023}. 
Depending on the sublimation timescale of the grains, the stream of dusty particulates may be sculpted into comet-like tails under the influence of gravity and radiation pressure from the host star (see, e.g., \citealt{ Brogi_2012, Budaj_2013, vanLieshout_2014, Sanchis-Ojeda_2015}). The blocking of starlight by the planet's dusty tails results in periodic transit profiles that lack the usual mirror symmetry around the time of minimum light. With continuous replenishment by time-varying outflows of surface material, these dusty tails also produce variable depths from transit to transit \citep{Rappaport_2014, Bromley_2023}. Asymmetric transit profiles and time-variable transit depths are the hallmark observations of a disintegrating or `catastrophically evaporating' planet (see also \citealt{vanLieshout_2018}, and references therein).

Three disintegrating planets have been discovered to date: \textit{Kepler}-1520\,b \citep{Rappaport_2012}, KOI-2700\,b \citep{Rappaport_2014}, and \textit{K2}-22\,b \citep{Sanchis-Ojeda_2015}, all of which were discovered using data from the \textit{Kepler} space telescope. These are classified as ultra-short period (USP) planets that orbit cool dwarf stars with spectral types ranging from mid-K to early-M. Based on the dust models that are discussed in this work and in previous studies (e.g., \citealt{Perez-Becker_2013}), these planets are inferred to have sizes significantly smaller than Earth and would have otherwise been undetectable in the photometric data if not for the distinct transits characteristic of their large, dusty tails. Although generally asymmetric, the transit profiles of each planet's dust tail are different from one another, reflecting the diversity of dust grain properties and dynamics unique to each system (e.g., \citealt{vanLieshout_2014}). Their transit depths are typically 0.5\% or smaller, and modeling suggests dust-driven mass loss at a rate that ranges from several lunar masses to one Earth mass per Gyr, such that the disintegration lifetimes of these planets are expected to be up to several hundred Myr depending on their ages and initial masses \citep{Perez-Becker_2013}. 

Studying the properties and the evolution of disintegrating rocky planets, however, has been challenging. The three known cases have relatively faint host stars ($V\apprge16$), which hinders follow-up observations with existing facilities (e.g., \citealt{Croll_2014, Bochinski_2015, Colon_2018, Gaidos_2019, Ridden-Harper_2019}). 
As a result, the hope moving forward has been to find more disintegrating exoplanets, particular those around brighter host stars. In this regard, the estimated occurrence rates of these objects, based on the models of \citet{Perez-Becker_2013} and \citet{Curry_2024}, suggest that there should be many more to be found \textemdash{} more so than were actually discovered in the \textit{Kepler} and \textit{K2} missions \citep{Borucki_2010, Howell_2014}.



NASA's Transiting Exoplanet Survey Satellite \citep[TESS;][]{Ricker_2015} has opened new prospects for discovering additional disintegrating planets. With follow-up observations as the primary consideration for its mission, TESS focuses on transiting planets orbiting bright stars \citep{Winn_2024}. Its nearly all-sky survey, which is expected to achieve $\sim$97\% coverage by the end of its ongoing second Extended Mission, provides vastly more opportunities for finding and characterizing disintegrating planets across a wider range of planet hosts. For the same level of photometric precision, a star observed by TESS typically appears $\sim4.5$ magnitudes fainter in the \textit{Kepler} bandpass (e.g., \citealt{Jenkins_2010,  Sullivan_2015, Kunimoto_2022_yield}). The previous three discoveries of disintegrating planets have \textit{Kepler} magnitudes between $15-16$. These correspond to TESS magnitudes of $10.5-11.5$, at which photometric precision levels with TESS are sufficient to detect transits with depths of $0.5-1.0\%$ \citep[Fig. 6]{Winn_2024}, which are characteristic of the previous three disintegrating planets. The major advantage with TESS observations, however, lies in its sample size, which provides far more opportunities to discover additional disintegrating planets. In particular, there are a total of $\sim2.5$ million unique stars brighter than a TESS magnitude of 11.5, which is a quantity several times larger in number than the sample size observed by \textit{Kepler} and \textit{K2} combined.


Here, we present the discovery of \system\,Ab, a fourth disintegrating planet. Discovered by TESS, the highly asymmetric transit light curves show evidence for \textit{both} leading and trailing dust tails, with variable transit depths typically greater than $1\%$. Remarkably, the planet is in a 30.5-hr orbit around a mid-K-type dwarf, making it the longest-period planet known to be disintegrating. With a TESS magnitude of 9.18, it is more amenable than the other three known disintegrating planets to detailed follow-up observations. A preview of the feasibility of such observations is presented in this study through our characterization of the host star as well as our preliminary modeling and interpretation of the dust tails based on the TESS observations.

\section{TESS Data} \label{section:data}

\system\,A (TIC~466376085, HIP~107587) was observed by TESS during Sector 55 (UT 2022 August 5 to 2022 September 1st) in the field of view of CCD 1 of Camera 1 and during Sector 82 (UT 2024 August 10 to 2024 September 5) in the field of view of CCD 4 of Camera 1. In both sectors, \system\,A data were obtained from the TESS Full Frame Images (FFIs), with a 10-minute (600 seconds) integration time during Sector 55 and a 200-second integration time in Sector 82. The FFI light curves of \system\,A were extracted using the TESS Quick-Look Pipeline (QLP; \citealt{Huang_2020a, Huang_2020b, Kunimoto_2022}), and can be found on MAST: \dataset[10.17909/t9-r086-e880]{http://dx.doi.org/10.17909/t9-r086-e880}.
A transit signal was detected initially as part of the QLP transit search of Sector 82. The search, which automatically includes all available TESS data for each target, detected transits at a 1.271869-day period using the Box Least Squares algorithm (BLS, \citealt{Kovacs_2002}; see also \citealt{Kunimoto_2023}). Subsequent visual vetting of the transits identified their highly asymmetric morphology \textemdash{} a rounded, short ingress followed by a long egress \textemdash{} as well as strong depth variations from transit to transit (\figr{fig1}a). We note that the transits of \system\,Ab were also detected in the Sector 55 TESS-SPOC FFI light curve \citep{TESSSPOC2020} using the Science Processing Operations Center pipeline \citep{jenkinsSPOC2016} with a noise-compensating matched filter \citep{2010SPIE.7740E..0DJ, 2020TPSkdph}, from which the pipeline's difference image test located the host star within $0.375\pm2.5^{\prime\prime}$ of the transit source \citep{Twicken:DVdiagnostics2018}.

To study these transits in further detail, we manually de-trended the FFI light curves extracted using the QLP's simple aperture photometry, one Sector at a time. Doing this required removing the slow trend underlying the light curve, which was performed as follows. We masked out the segments of the light curve containing transits. We defined the zero point of each transit's orbital phase ($t_0$) to be the time of minimum flux. Next, we masked out in-transit segments of the light curve, which are timestamps with phases $\geq$$-0.25$ or $<$$\,0.45$ (see \figr{fig1}b). We designated the median flux and timestamp of each continuous segment of the masked light curve as interpolating points in a Piecewise Cubic Hermite Interpolating Polynomial \citep{Fritsch_1984}. The full, unmasked light curve was then divided by the fitted polynomial to remove underlying slow trends and to set the out-of-transit flux levels equal to unity.

We obtained an average transit profile by phase-folding the de-trended light curve and binning over 6-minute intervals. This yielded a smoothed phase-folded light curve comprising 303 binned flux values with phases $\in[-0.5, 0.5]$ for modeling the average transit properties (see Section \ref{section:light_curve}). The average transit profile in Fig. \ref{fig:fig1}b-c reveals a total duration of about 20 hours (65\% of the orbital period) from the beginning of the ingress until the time at which out-of-transit flux is regained. The main contributor to the transit duration comes from the egress, implying a particularly long trailing dust tail.


\section{Ground-based Observations} \label{section:aux_obs}

\subsection{ASAS-SN Photometry} \label{section:asassn}

\begin{figure}
    \centering
	\includegraphics[width=1.\columnwidth]{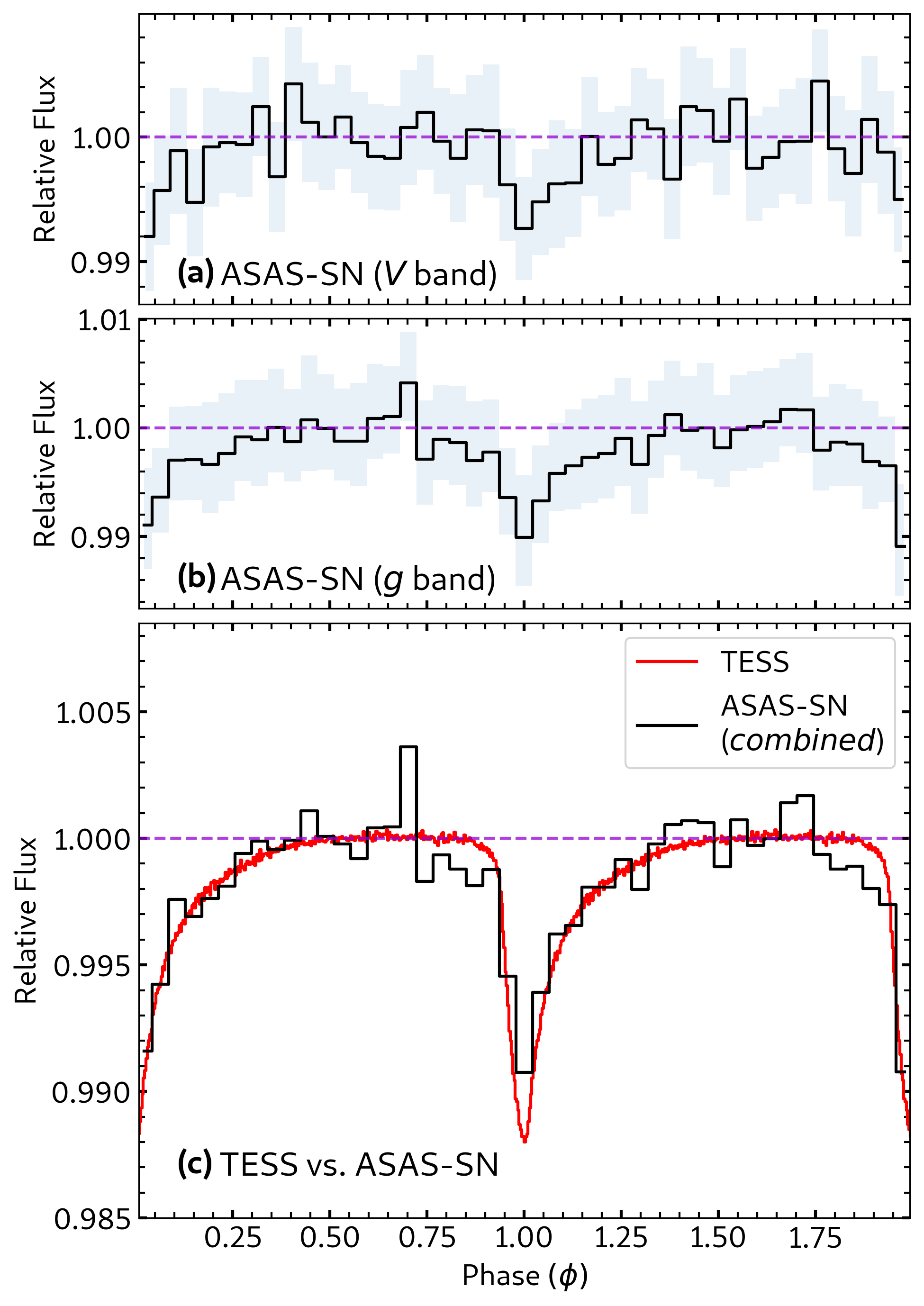}
    \caption{(a-b) Light curves of \system\,A from the All-Sky Automated Survey for Supernovae (ASAS-SN) folded at a period of 1.271869 days at the ephemerides determined from TESS data. The ASAS-SN light curves are binned over 75-minute intervals. The shaded bands indicate the formal and binned flux uncertainties added in quadrature. (c) A comparison between the TESS and ASAS-SN phase-folded light curves. }
    \label{fig:fig2}
\end{figure}

To verify the authenticity and persistence of the transits discovered in the TESS data, we examined ground-based time series photometry of \system\,A collected by the All-Sky Automated Survey for Supernovae (ASAS-SN; \citealt{Shappee_2014}). The ASAS-SN data comprises 3,333 measurements (3,155 with good quality flags) in the $g$ band, and 1,064 measurements (1,044 with good quality flags) in the $V$ band, obtained between 2012 October 21 and 2024 September 23. A BLS transform of the ASAS-SN data was computed. The largest peak between periods of 1 and 10 days is at $P = 1.271884$ days, which has an uncertainty of about one part in $10^5$ but differs from the TESS period by only one part in 83,000. To demonstrate the consistency of these period measurements, we present in \figr{fig2} the $5$-$\sigma$-clipped ASAS-SN data folded on the TESS period, where we find a similar signal to that seen in the TESS data, but with a lower signal-to-noise ratio as expected. Remarkably, the transit depth, when averaged over more than a decade, is roughly consistent with that observed by TESS.  This demonstrates that the dusty transits have persisted for at least the past decade.

\subsection{Las Cumbres Observatory Global Telescope (LCOGT) Photometry} \label{section:lco}

\begin{figure}
    \centering
	\includegraphics[width=1.\columnwidth]{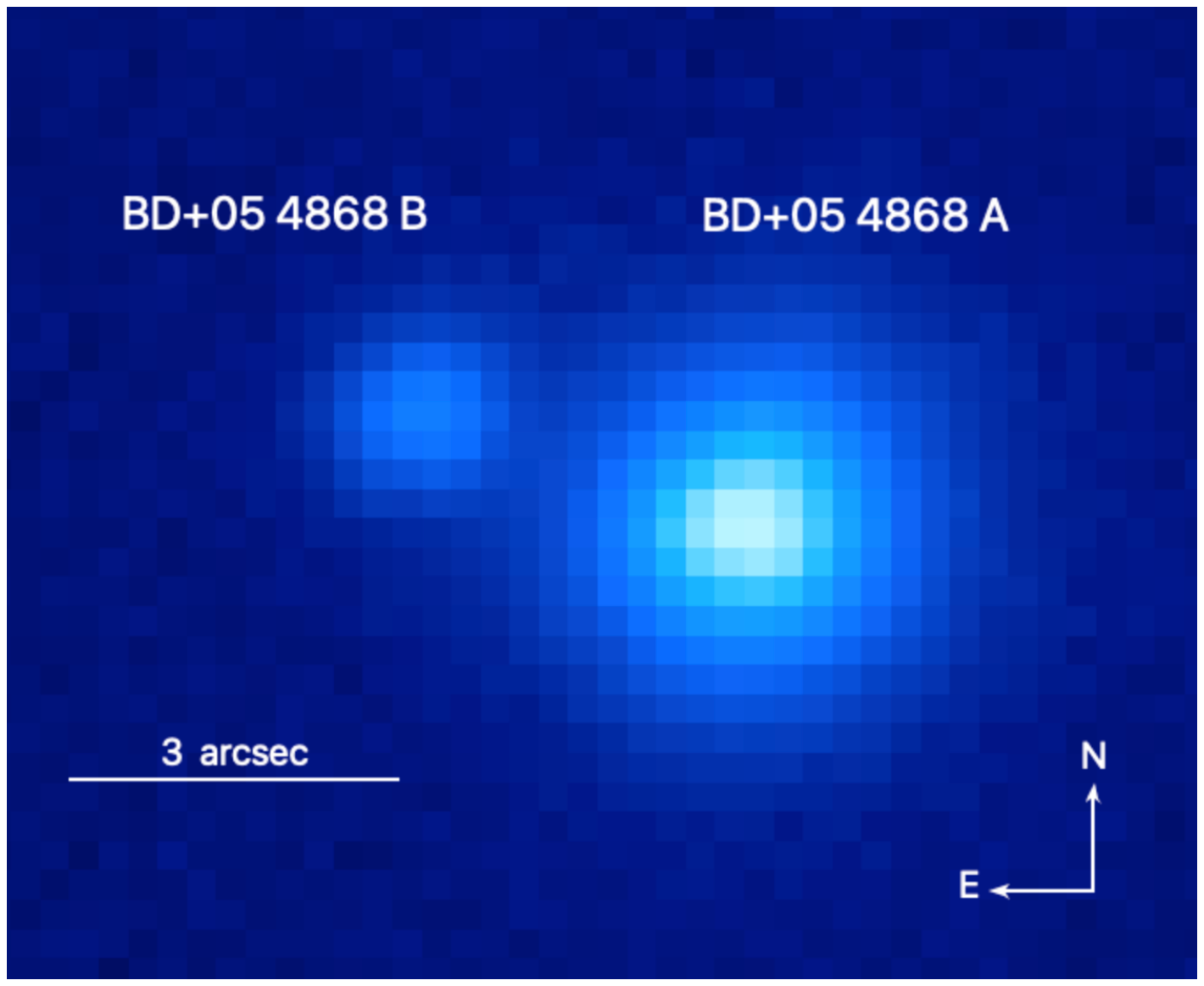}
    \caption{A zoomed-in view of the target, \system\,A, and its fainter binary companion, \system\,B, located 3.0$^{\prime\prime}$ to the North-East. This image is taken from one of the $r_p$-band frames obtained with LCO 2.0m/MuSCAT3 on the night of UT 2024 October 17. It shows that the two binary components were resolved in this observation. \figr{lco} shows the light curves extracted from the MuSCAT3 data obtained that night.}
    \label{fig:zoom-in}
\end{figure}

The TESS pixel scale is $21\arcsec$ pixel$^{-1}$ \citep{Ricker_2015} and photometric apertures typically extend out to roughly 1 arcminute, generally causing multiple stars to blend in the TESS aperture. To determine the true source of transit signals in the TESS data and improve the transit ephemerides,
we conducted ground-based lightcurve follow-up observations of the field around \system\,A as part of the \textit{TESS} Follow-up Observing Program\footnote{https://tess.mit.edu/followup} Sub Group 1 \citep[TFOP;][]{collins:2019}. The target has a comoving visual binary companion (\system\,B, TIC 2025181668, see \figr{zoom-in}) located 3.0$^{\prime\prime}$ to the North-East of the target that is 3.1 mag fainter than the target in the TESS band and 2.9 mag fainter in the $G$ band. We used the {\tt TESS Transit Finder}, which is a customized version of the {\tt Tapir} software package \citep{Jensen:2013}, to schedule our transit observations. 

We observed a full transit window of \system\,A on UT 2024 October 17 using the Las Cumbres Observatory Global Telescope \citep[LCOGT;][]{Brown_2013} 2\,m Faulkes Telescope North at Haleakala Observatory on Maui, Hawai'i. The telescope is equipped with the MuSCAT3 simultaneous multi-band imager \citep{Narita:2020} that has a field of view of 9.1$^{\prime}$ $\times$ 9.1$^{\prime}$ and a pixel scale of 0.265$^{\prime\prime}$ per pixel. Observations were made in-focus in order to resolve the binary companion, and the exposure times were 3, 1, 1, and 2 seconds in the $g_p$, $r_p$, $i_p$, and $z_s$ bands, respectively. All images were calibrated by the standard LCOGT {\tt BANZAI} pipeline \citep{McCully:2018} and differential photometric data were extracted using {\tt AstroImageJ} \citep{Collins:2017}.  We used circular photometric apertures with radius $1\farcs9$ that excluded most of the flux from the visual binary companion. 

\begin{figure}
    \centering
	\includegraphics[width=1.\columnwidth]{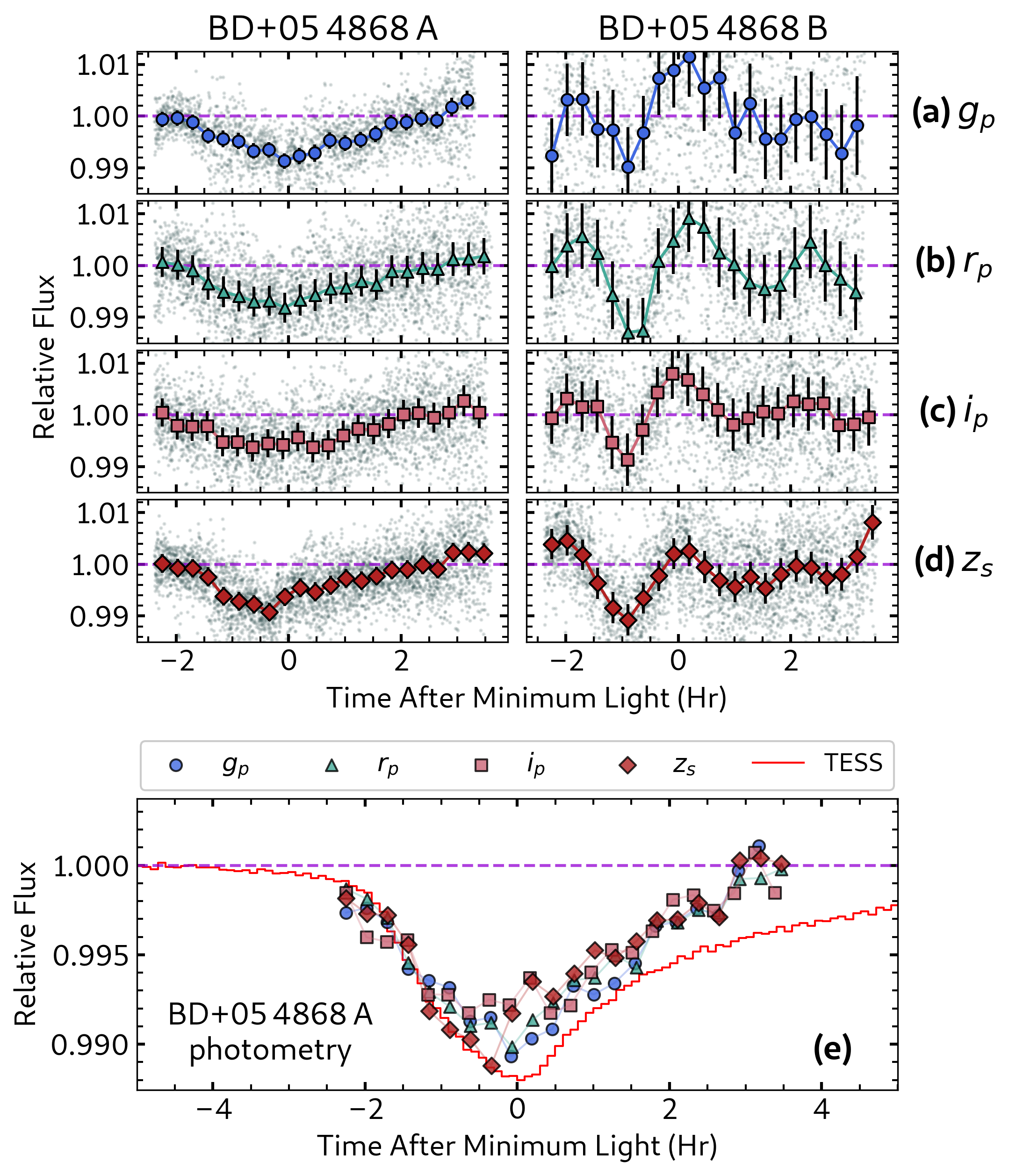}
    \caption{LCO 2.0m/MuSCAT3 light curves of the target, \system\,A, and its fainter binary companion, \system\,B, obtained on UT 2024 October 17. (a) The light curves of each of the four bands are shown in the top panels, with the data in small gray points overplotted by binned light curves. The $g_p$, $r_p$, $i_p$, and $z_s$ binned light curves are plotted in blue circles, green triangles, orange squares, and red diamonds, respectively.
    The \system\,A light curve is also shown in the bottom panel, indicating that it is similar in depth and overall shape to that of the TESS light curves (red solid line), although there is a small difference due the varying shape of the transits across each orbit (see \figr{montage}).
    (e) A comparison of transits in the LCO light curve with the observed mean transit profile from TESS data. A small flux offset was applied to the LCO light curves in this panel to align them with the TESS transit profile.}
    \label{fig:lco}
\end{figure}

\figr{zoom-in} shows one of the MuSCAT3 images in the $r_p$ band, demonstrating that the two stars were well resolved. \figr{lco} shows the target light curve in each of the four bands, alongside the light curve of the visual binary companion \system\,B. The target light curve shows a similar shape to that of the TESS and ASAS-SN phase folded light curve, including a similar transit depth. If \system\,B were the origin of the transit signal, variability at the 10\% level would be expected as a result of blending with \system\,A in the TESS pixels and the brightness difference between the two stars. The light curve of the companion, however, does \emph{not} show such variability, confirming that the transit signal originates from the primary in the visual binary system.

\subsection{High Angular Resolution Imaging} \label{section:imaging}
To further rule out possible contamination by closely bound or line-of-sight companions, we observed \system\,A with NIRC2 adaptive optics (AO) imaging at the Keck Observatory. Observations were made on UT 2024 September 8 behind the natural guide star systems \citep{dekany2013,wizinowich2000} in the narrow band J continuum (\texttt{Jcont}; $\lambda_o = 1.2132; \Delta\lambda = 0.0198~\mu$m) and the narrow band K continuum (\texttt{Kcont}; $\lambda_o = 2.2706; \Delta\lambda = 0.0296~\mu$m) filters. The narrow angle mode was used, providing a pixel scale of around 0.01\arcsec~px$^{-1}$ and a full field of view of about 10\arcsec. A standard three-point dither pattern was used to avoid the lower-left quadrant of the detector, which is typically noisier than the other three quadrants. The dither pattern has a step size of 3\arcsec. Each dither position was observed three times, with 0.5\arcsec~positional offsets between each observation, for a total of nine frames. The reduced science frames were combined into a single image with final resolutions of 0.04\arcsec\ and 0.05\arcsec\ at \texttt{Jcont} and \texttt{Kcont}, respectively.

\begin{figure}
    \centering
	\includegraphics[width=1.\columnwidth]{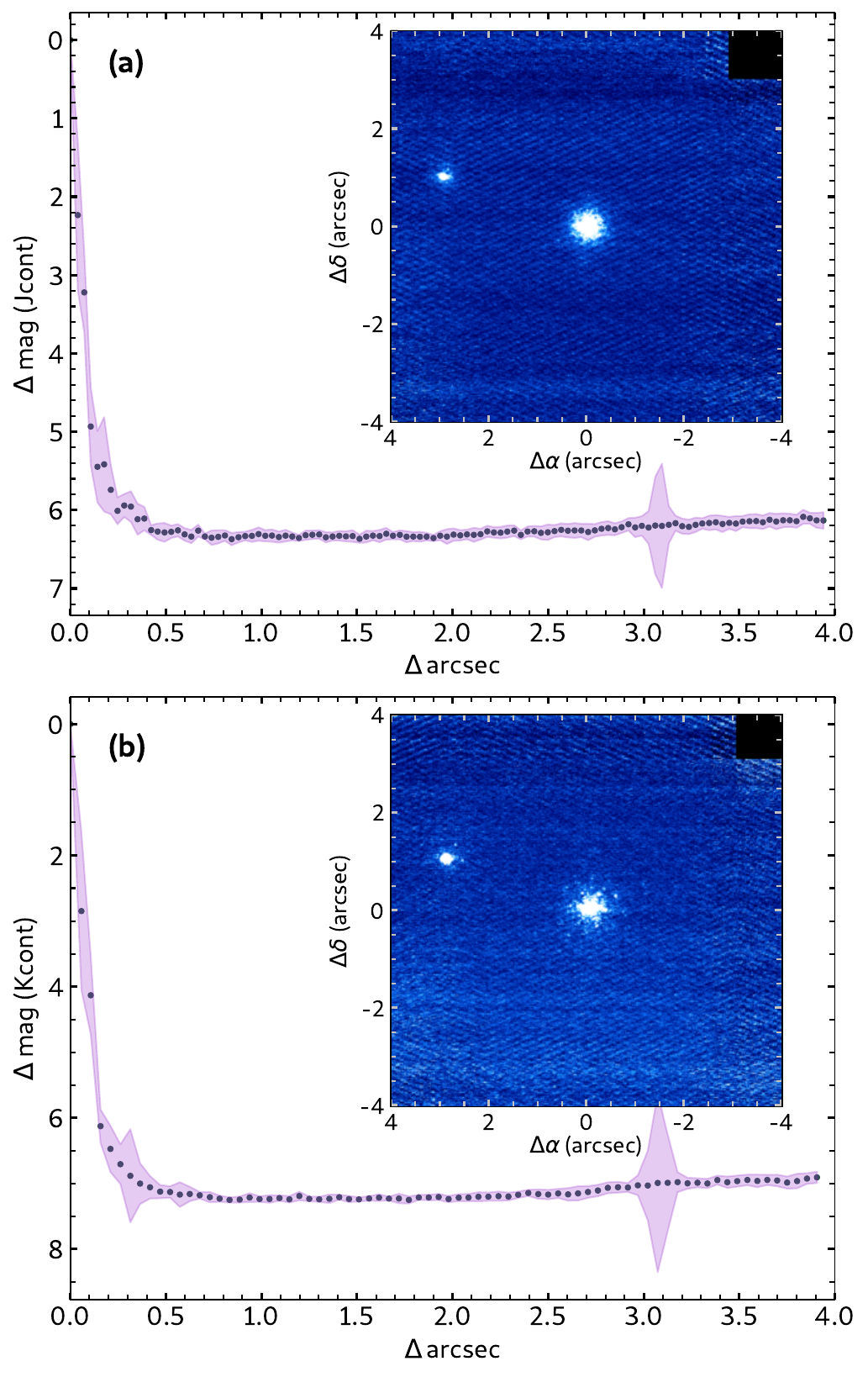}
    \caption{Near-infrared adaptive optics imaging and corresponding sensitivity curves for \system\,A in (a) the narrow band J continuum (\texttt{Jcont}) and in (b) the narrow band K continuum (\texttt{Kcont}). The increased uncertainty at  $\Delta\text{arcsec}\sim3.0^{\prime\prime}$ is caused by the presence of \system\,B. }
    \label{fig:keck_ao}
\end{figure}

The sensitivity of the final combined AO image was determined by injecting simulated sources azimuthally around the primary target every $20^\circ $ at separations of integer multiples of the central source's FWHM \citep{furlan2017}. The brightness of each injected source was scaled until standard aperture photometry detected it with 5-$\sigma $ significance.  The final 5-$\sigma $ limit at each separation was determined from the average of all of the determined limits at that separation and the uncertainty on the limit was set by the root-mean-squared dispersion of the azimuthal slices at a given radial distance. The final images and sensitivities are shown in Figure~\ref{fig:keck_ao}. Other than \system\,B, no other nearby stellar companions are identified within our detection limits.

\section{Host Star Characterization}

\subsection{NEID Spectroscopy} \label{section:neid}

We obtained seven spectra of \system\,A with the NEID spectrograph \citep{Schwab_2016} on the WIYN 3.5m Telescope at Kitt Peak Observatory. The spectra were collected in order to look for radial velocity (RV) variability and to measure the host star's spectral parameters.

NEID is a high-resolution spectrograph that covers a wide wavelength range from 380 to 930 nm. We observed in high resolution (HR) mode (R$\sim$120,000) with exposure times of 600 seconds, resulting in a signal-to-noise ratio of 42 -- 66 per pixel at 5500 \AA.
The spectra were processed with the NEID Data Reduction Pipeline (DRP)\footnote{\href{https://neid.ipac.caltech.edu/docs/NEID-DRP/}{https://neid.ipac.caltech.edu/docs/NEID-DRP/}}. The NEID RVs are listed in \tabr{NEID_RVs}.
The seven RVs do not show any variability exceeding a few m\,s$^{-1}$, with a root-mean-square of 2.3~m\,s$^{-1}$ and individual RV errors in the range of 1.3-2.3~m\,s$^{-1}$. Through the injection and recovery of simulated RV signals of a circular orbit with a period and phase matching the transit ephemerides, we place a 5-$\sigma$ upper limit of 4.5~m\,s$^{-1}$ on the RV semi-amplitude, which corresponds to a mass of 6.2 M$_\oplus$ given the host star's mass and the orbital period. Therefore, the transiting object must have a mass of a small terrestrial planet or smaller. More generally, we fitted assumed circular orbits with periods of $0.5-1,000$\,days to the RV data to identify the maximum mass of a planet that may remain undetected in the system from our observations, as described in Appendix \ref{appendix:rv_lim}. We find that no Jupiter-mass planets with periods $<100\,$d may remain undetected, while brown dwarfs are ruled out for periods shorter than a year.

To determine the stellar spectroscopic parameters, we blaze-corrected the spectra using the NEID blaze function, shifted them to the solar rest frame according to the barycentric correction and RVs from the NEID DRP, and then stacked them using a signal-to-noise ratio weighted average. We applied the code \texttt{SpecMatch-Emp} \citep{Yee_2017} to the stacked spectrum to identify the fundamental parameters of \system\,A through a search for matches across a library of empirical stellar spectra. The code retrieved a good match corresponding to a K-dwarf with $T_{\mathrm{eff}}=4540\pm110\,$K, $R_{\star}=0.75\pm0.08R_\odot$, $\log(g)=4.58\pm0.12\,$dex, and [Fe/H] = $-0.07\pm0.09\,$dex. In addition, we used \texttt{SpecMatch-Syn} \citep{Petigura_2015} to match the spectra to the \citet{Coelho_2005} synthetic spectral library, incorporating a broadening kernel that includes contributions from the intrinsic instrumental line profile (measured empirically from telluric lines), macroturbulent broadening assuming the \citet{Valenti_2005} relation, and rotational broadening. The algorithm determined $v\sin i = 1.9 \pm 1.0\,$km s$^{-1}$, which we interpret as an upper limit of $3\,$km\,s$^{-1}$ following \cite{Masuda_2022}. The NEID DRP additionally extracts activity indices in the form of Ca II H\&K S-indices (\citealt{Duncan_1991}, tabulated in Table \ref{table:NEID_RVs}), from which we identified no statistically significant correlation with the RVs.

\begin{table}[ht]
    \centering
    \caption{NEID radial velocity (RV) measurements and the associated Ca II H\&K S-indices as output by the NEID Data Reduction Pipeline. $\sigma_{\mathrm{RV}}$ and $\sigma_{\mathrm{S-index}}$ correspond to the uncertainties of the RV and S-index measurements, respectively.}\label{table:NEID_RVs}
    \begin{tabular}{ccccc}
    \hline 
    Time & RV & $\sigma_{\mathrm{RV}}$& S-index & $\sigma_{\mathrm{S-index}}$\\
    BJD & m s$^{-1}$ & m s$^{-1}$ & & \\
    \hline 
    2460605.635696 &-25,285.0 &1.8 & 0.34 & 0.01\\
    2460606.588164 &-25,288.1 &1.4 & 0.36 & 0.01\\
    2460607.577608 &-25,289.9 &2.3 & 0.36 & 0.02\\
    2460608.583141 &-25,286.1 &1.5 & 0.36 & 0.01\\
    2460611.581506 &-25,285.7 &1.3 & 0.36 & 0.01\\
    2460615.609390 &-25,286.0 &1.5 & 0.35 & 0.01\\
    2460619.610952 &-25,290.9 &2.0 & 0.33 & 0.01\\
    \hline 
    \end{tabular} 
\end{table}  

\begin{figure}
    \centering
	\includegraphics[width=1.\columnwidth]{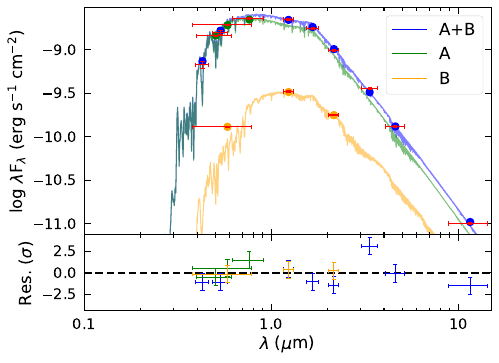}
    \caption{Spectral energy distribution (SED) of \system\,AB. The crosses show the observed broadband photometry, with horizontal widths indicating the filter bandwidths. The circles represent the model fluxes in the corresponding bands. Colors denote whether the observation corresponds to the primary star (green), secondary star (yellow), or the blended pair (blue). The bottom panel shows the deviations between the data and the model divided by the photometric uncertainties.
    For illustrative purposes, we also show the \citet{Kurucz_1993} model atmospheres for the best-fitting stellar models. These models were not used directly in the fit, although they were used to generate the MIST bolometric correction tables.}
    \label{fig:sed}
\end{figure}

\subsection{Binary Star Spectral Energy Distribution (SED)} \label{section:sed}

We modeled the spectral energy distribution of the \system\,AB binary system using \texttt{Exofastv2} \citep{ExoFASTv2_Eastman19}, which fits observed broadband photometry to bolometric correction tables derived for the MIST grid of stellar models \citep{Choi_2016}. We used photometry from the Tycho \citep{Hog_2000}, Gaia DR3 \citep{Gaia_DR3}, 2MASS \citep{Cutri_2003,Skrutskie_2006}, and WISE catalogs \citep{Cutri_2012}. Given the 3$^{\prime\prime}$ separation of the binary system, we assumed that the Tycho, 2MASS, and WISE photometry were blended. Gaia resolved the pair and provided $G$-band magnitudes for the two stars individually, but reported only $G_\mathrm{BP}$ and $G_\mathrm{RP}$ photometry for the primary. We imposed error floors of 0.02~mag on the Gaia photometry to account for systematic uncertainties in the absolute calibration of model photometry. We also made use of the differential magnitudes in \texttt{Jcont} and \texttt{Kcont} obtained from the AO imaging in Section \ref{section:imaging}.

In modeling the two stars, we imposed a prior on the primary star's [Fe/H] based on our spectroscopic analysis (\S\ref{section:neid}), as well as a prior on both stars' distances from their respective Gaia parallaxes, corrected for the known zero-point error \citep{Gaia_DR3}. We assumed no interstellar extinction ($A_V=0$), given the proximity of the system at $\approx44\,$pc. We required the two stellar models to have the same age and initial metallicity.

The resulting fit is shown in Fig. \ref{fig:sed}, with the fitted parameters tabulated in Table \ref{table:properties}. The mass of the primary \textemdash{} the planet host \textemdash{} is $M_\star=0.70\pm0.02M_{\odot}$, with $T_\mathrm{eff}$, $R_\star$, and [Fe/H] fully consistent with the spectroscopically-derived parameters. Its near solar-metallicity and sub-solar mass, combined with the lack of observed stellar rotation and activity characteristic of young stars, suggests an age comparable or older to that of the Sun. Hereafter, we adopt the stellar parameters derived from SED fitting as the definitive set for our analyses of the host star. 

The secondary star \system\,B was determined by the SED fitting to have a mass of $0.43\pm0.03\,M_\odot$ and $T_\mathrm{eff} = 3480\pm70$~K, consistent with the properties of an M-dwarf. The similarity of the two stars' proper motions and radial velocities in Table \ref{table:properties} are consistent with their nature of being gravitationally bound to one another, with a projected physical separation of $132.037\pm0.002\,$AU. To get an estimate of a 3D separation between both stars, we fitted orbital parameters to their \textit{Gaia} DR3 astrometry using Orbits For The Impatient (OFTI, \citealt{Blunt_2017}) as implemented by \citet{Pearce_2020}. A semi-major axis of $185^{+97}_{-61}$ AU is reported\footnote{The reported uncertainties are the 16th and 84th percentiles of the marginal distribution.} from the posterior distribution of of 10,000 accepted orbital fits, which corresponds to a binary orbital period of $\sim$1,000-4,000 years. We additionally adopted the approach by \citet[their Section 5.2]{Zasche_2023}, which adopts a Monte Carlo approach to find semi-major axes that solves the energy equation for the orbit. We found that the range of separations yielded by this alternative approach are consistent with those estimated by OFTI. We assert, however, that these results on the binary system's orbit are indicative only, given that \system\,B's high RUWE value of 1.5 may suggest a poor fit to its existing astrometric measurements.

\begin{table}[ht]
    \centering
    \caption{Properties of the \system\,AB system and the planet \system\,Ab. The adopted astrometric measurements are from \textit{Gaia} DR3 \citep{Gaia_DR3}. Numerical quantities in parentheses represent uncertainties in the last decimal place.}\label{table:properties}
    \begin{tabular}{cccccc}
        \toprule
        \multicolumn{6}{c}{\textbf{\system\,A Properties}} \\ 
        \midrule
        \multicolumn{2}{c}{TIC ID} &  \multicolumn{4}{c}{466376085} \\
        \multicolumn{2}{c}{\textit{Gaia} DR3 Source ID} &  \multicolumn{4}{c}{2700378125204437760} \\
        \multicolumn{2}{c}{\textit{Gaia} $G$-band Magnitude} & \multicolumn{4}{c}{9.84} \\
        \multicolumn{2}{c}{2MASS K-band Magnitude} & \multicolumn{4}{c}{7.45} \\
        \multicolumn{2}{c}{Distance (pc)}  & \multicolumn{4}{c}{$43.67\pm0.03$} \\
        \multicolumn{2}{c}{$\mu_{\alpha^\star}$ (mas yr$^{-1}$)}  & \multicolumn{4}{c}{$205.60\pm0.02$} \\
        \multicolumn{2}{c}{$\mu_{\delta}$ (mas yr$^{-1}$)}  & \multicolumn{4}{c}{$106.36\pm0.01$} \\
        \multicolumn{2}{c}{\textit{Gaia} DR3 Radial Velocity (km s$^{-1}$)} & \multicolumn{4}{c}{$-25.6\pm0.2$} \\
        \multicolumn{2}{c}{$T_{\mathrm{eff, spect}}$ (K)} & \multicolumn{4}{c}{\hspace{-2.5pt}$4540\pm110$} \\
        \multicolumn{2}{c}{$T_{\mathrm{eff, SED}}$ (K)} & \multicolumn{4}{c}{\hspace{-2.5pt}$4596^{+65}_{-64}$} \\
        \multicolumn{2}{c}{$[$Fe/H$]_{\mathrm{spect}}$ (dex)} & \multicolumn{4}{c}{\hspace{-1.5pt}$-0.07\pm0.09$} \\
        \multicolumn{2}{c}{$[$Fe/H$]_{\mathrm{SED}}$ (dex)} & \multicolumn{4}{c}{$-0.05^{+0.08}_{-0.05}$} \\
        \multicolumn{2}{c}{$\log (g)_{\mathrm{spect}}$ (dex)} & \multicolumn{4}{c}{\hspace{-1.5pt}$4.58\pm0.12$} \\
        \multicolumn{2}{c}{$\log (g)_{\mathrm{SED}}$ (dex)} & \multicolumn{4}{c}{\hspace{-1.5pt}$4.60^{+0.03}_{-0.02}$} \\
        \multicolumn{2}{c}{$v\sin i$ (km$\,$s$^{-1}$)} & \multicolumn{4}{c}{$<3$} \\
        \multicolumn{2}{c}{Mass, $M_{\star}$ ($M_{\odot}$)} & \multicolumn{4}{c}{$0.70\pm0.02$}\\
        \multicolumn{2}{c}{Radius, $R_{\star}$($R_{\odot}$)} & \multicolumn{4}{c}{$0.69\pm0.02$}\\
        \multicolumn{2}{c}{Luminosity, $L_{\star}$ ($L_{\odot}$)} & \multicolumn{4}{c}{$0.192\pm0.005$}\\
        \multicolumn{2}{c}{Age (Gyr)} & \multicolumn{4}{c}{$11.1^{+1.7}_{-3.0}$}\\
        \midrule
        \multicolumn{6}{c}{\textbf{\system\,B Properties}} \\ 
        \midrule
        \multicolumn{2}{c}{TIC ID} &  \multicolumn{4}{c}{2025181668} \\
        \multicolumn{2}{c}{\textit{Gaia} DR3 Source ID} &  \multicolumn{4}{c}{2700378125203895808} \\
        \multicolumn{2}{c}{\textit{Gaia} $G$-band Magnitude} & \multicolumn{4}{c}{12.77} \\
        \multicolumn{2}{c}{Distance (pc)}  & \multicolumn{4}{c}{$43.2\pm0.12$} \\
        \multicolumn{2}{c}{$\mu_{\alpha^\star}$ (mas yr$^{-1}$)}  & \multicolumn{4}{c}{$212.16\pm0.08$} \\
        \multicolumn{2}{c}{$\mu_{\delta}$ (mas yr$^{-1}$)}  & \multicolumn{4}{c}{$112.17\pm0.04$} \\
        \multicolumn{2}{c}{\textit{Gaia} DR3 Radial Velocity (km s$^{-1}$)} & \multicolumn{4}{c}{$-24.6\pm1.6$} \\
        \multicolumn{2}{c}{$\log (g)_{\mathrm{SED}}$ (dex)} & \multicolumn{4}{c}{\hspace{-1.5pt}$4.83\pm0.03$} \\
        \multicolumn{2}{c}{Mass, $M_{\star}$ ($M_{\odot}$)} & \multicolumn{4}{c}{$0.43\pm0.03$}\\
        \multicolumn{2}{c}{Radius, $R_{\star}$($R_{\odot}$)} & \multicolumn{4}{c}{$0.42\pm0.02$}\\
        \multicolumn{2}{c}{Luminosity, $L_{\star}$ ($L_{\odot}$)} & \multicolumn{4}{c}{$0.023\pm0.001$}\\
        \midrule
        \multicolumn{6}{c}{\textbf{\system\,Ab Properties}} \\ 
        \midrule
        \multicolumn{2}{c}{Period, $P$ ($d$)} &  \multicolumn{4}{c}{1.271869$\,$(1)} \\
        \multicolumn{2}{c}{Transit Epoch, $t_0$ (BJD$-$2457000)} & \multicolumn{4}{c}{3556.2892$\,$(3)} \\
        \multicolumn{2}{c}{Transit Depth, $\delta$} & \multicolumn{4}{c}{$0.8-2.0\%$}  \\
        \multicolumn{2}{c}{Mean Transit Depth, $\langle\delta\rangle$} & \multicolumn{4}{c}{$1.2\%$}  \\
        \multicolumn{2}{c}{Semi-major Axis, $a$ (AU)} & \multicolumn{4}{c}{0.0208$\,$(3)} \\
        \multicolumn{2}{c}{Scaled Semi-major Axis, $a/R_{\star}$} & \multicolumn{4}{c}{6.4$\,\pm\,$0.2} \\
        \multicolumn{2}{c}{Sub-stellar Temperature, $T_{\mathrm{eq}}$ (K)\textsuperscript{\textdagger}} & \multicolumn{4}{c}{ $1820\pm45$} \\
        \bottomrule
        \textsuperscript{\textdagger} $T_{\mathrm{eq}}=T_{\mathrm{eff, SED}}\cdot\sqrt{R_{\star}/a}$
    \end{tabular}
\end{table}

\section{Light Curve Modeling}\label{section:light_curve}

\subsection{TESS Light Curves}
Models of dusty transit profiles (e.g., \citealt{Brogi_2012, Budaj_2013, vanLieshout_2014, vanWerkhoven_2014, vanLieshout_2016, Garai_2018, Baka_2021}) consider the distribution of extinction cross-section in the azimuthal direction of the planet's orbit to be the primary contributor to the transit shape. As described by \citet[their Appendix C.2]{Rappaport_2014}, the angular decay of this cross-section can be modeled in the form of an exponential distribution given a number of assumptions related to the dust's sublimation properties and dynamics (e.g., \citealt{vanLieshout_2014}). The exponential formulation, however, describes a decay over time, while a fit to the observed transits measures a decay over distance, which depends on both time and the properties of the dust grains. Therefore, we adopt a dust extinction profile, $\mathcal{Y}(d)$, that offers greater flexibility than a simple exponential function to model the dust profile as a function of its position, $x$, along the stellar disk:
\begin{equation}\label{eqn:mean_transit_fit}
    \mathcal{Y}(d) =
\begin{cases}
    E_{f}\,\exp[- (d/x_f)^g ], & \text{for } d > 0 \\
    E_{b}\,\exp[- (|d|/x_b)^g ], & \text{for } d \leq 0
\end{cases}
\end{equation}
where $d = x-x_p$ is the angular distance of a dust grain at $x$ from the angular location of the planet at $x_p$. These quantities are expressed in units of the host star's radius $R_{\star}$. The orbital phase of the planet at $x_p$ is taken to be zero when the planet crosses the central longitude of the host star\footnote{This is distinct from the phase of the observed transit, which is taken to be zero at minimum light.}.
Altogether, Eqn. \ref{eqn:mean_transit_fit} describes a superposition of profiles from the leading and trailing tails, which are modeled individually as Generalized Gaussian distributions sharing a common exponent $g$. These distributions are a generalization of the exponential distribution (i.e., $g=1$) and can represent steeper or shallower decay profiles. The Generalized Gaussian for the leading (trailing) tail profile is parameterized by its amplitude $E_f$ ($E_b$) and its scale length $x_f$ ($x_b$).


To generate a model of the mean transit profile, $\mathcal{T}(\phi)$, we integrate $\mathcal{Y}(d)$ over the dust tail where it overlaps with the stellar disk, and repeating the integration in increments of 6 minutes as the dust tails cross the disk. We apply quadratic limb-darkening coefficients to the stellar disk by linearly interpolating the \citet{Claret_2017} grid of coefficients\footnote{The grid of limb-darkening coefficients is based on the least square method applied to PHOENIX-COND solar-metallicity stellar atmosphere models \citep{Husser_2013}.} to the star's spectroscopic $T_{\mathrm{eff}}$ and $\log(g)$. In total, we fitted seven free parameters to the mean transit profile of \system\,Ab as tabulated in Table \ref{table:mean_transit_fit} in Appendix \ref{appendix:posterior}. The joint posterior distributions, which are also presented there, were sampled using nested sampling \citep{Higson_2019} in which light curves generated from each sampled set of parameters were evaluated by the $\chi^2$ value of the fit to the observed data.

\begin{figure}
    \centering
	\includegraphics[width=1.\columnwidth]{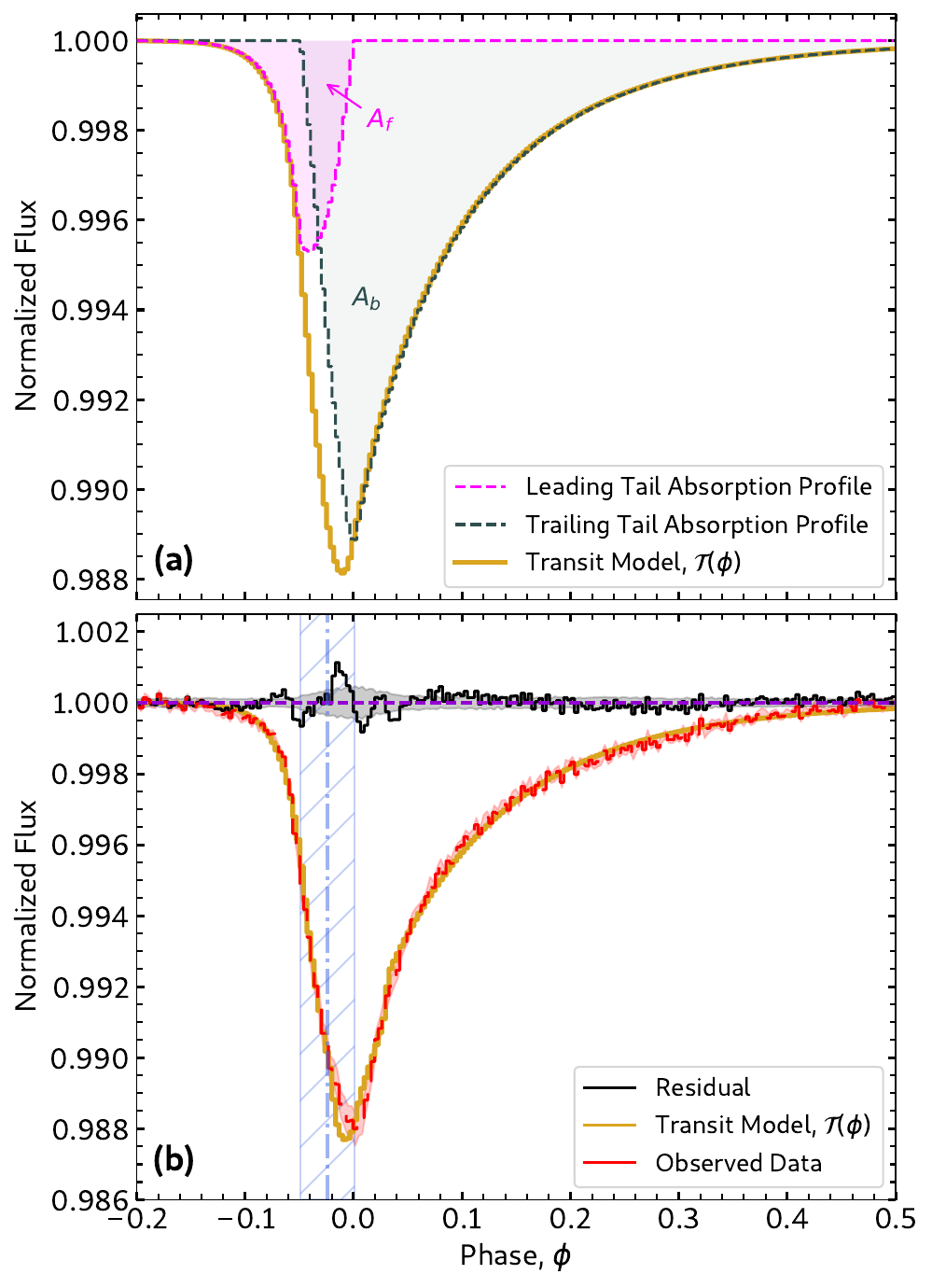}
    \caption{The fit to the mean transit profile of \system\,A using the transit model $\mathcal{T}(\phi)$, which convolves the dust extinction profile $\mathcal{Y}(d)$ (Eqn. \ref{eqn:mean_transit_fit}) with the stellar disk. (a) Decomposition of $\mathcal{Y}(d)$ as a superposition of two Generalized Gaussian distributions representing the trailing and leading tails of the dust cloud. The areas beneath the extinction profiles described by these distributions are indicated by $A_b$ and $A_f$, respectively. (b) Comparison of the model fit with the observed mean transit profile from TESS data (red). The blue vertical line at $\phi=-0.024$ indicates when the planet crosses the central longitude of the host star, with the hatched region indicating the star's diameter.}
    \label{fig:transit_model}
\end{figure}

The main quantities of interest from the fit are $x_b$ and $x_f$, the scale lengths of the trailing and leading tails, respectively. We estimated a ratio $x_b/x_f\simeq 6$, indicating a factor of six difference between the length of each tail. This is an important quantity that we will revisit in Section \ref{section:dust_dynamics} to constrain the dynamics of the dust tails. Meanwhile, the integrated area beneath each tail's attenuation profile (i.e., $A_b$ and $A_f$ in Fig. \ref{fig:transit_model}a) represents the collective extinction from the tail. Assuming that the optical properties of the grains are not highly disparate across both tails, and that the tails are not optically thick everywhere, the integrated area is indicative of the relative number density of grains within the tails (e.g., \citealt{vanLieshout_2014}), suggestive of a trailing tail more highly populated in dust grains relative to the leading tail. From our fitted model, we also find that the time of minimum light occurs at $x_p=0.97\,R_{\star}$, which is when the planet has almost traversed the entire stellar disk (see Fig \ref{fig:transit_model}b).

\begin{figure*}
    \centering
	\includegraphics[width=2.\columnwidth]{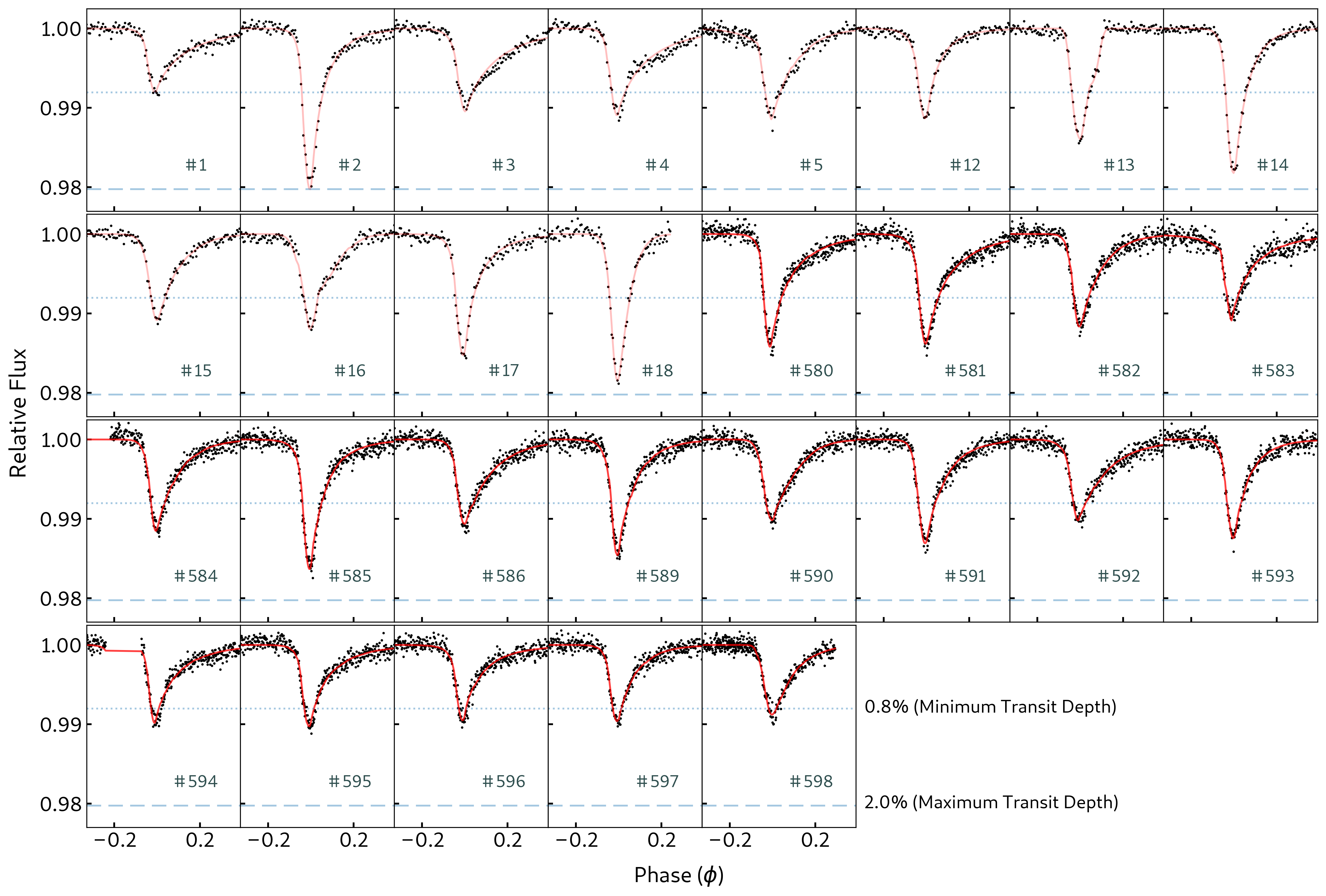}
    \caption{Individual transits of \system\,Ab from the TESS light curve. Transits are numbered based on the cycle count from the first transit at $t_{0, \#1}=\,2459798.2587$ (BJD), taking into account the presence of gaps in the light curve. Transits 1-18 are from TESS Sector 55, while Transits 580-598 are from TESS Sector 82. The mean transit depth across these transits is 1.2\%. Also indicated are the minimum $(0.8\%,$ dashed line) and maximum $(2.0\%,$ dotted line) transit depths across these observations. The red curve in each panel represents the model transit profile $\mathcal{T}(\phi)$ (Eqn. \ref{eqn:mean_transit_fit}) that is fit to each transit individually.}
    \label{fig:montage}
\end{figure*}

The high-cadence sampling of TESS data enables the detailed examination of individual transits from \system\,A's light curve, as shown in Fig. \ref{fig:montage}. We fit $\mathcal{T}(\phi)$ separately to each transit, with the resulting measurements presented in Appendix \ref{appendix:individual_transits}. 
The variable transit depths are thought to arise from a limit cycle involving stellar insolation and mass loss that switches between low and high dust emission rates \citep{Rappaport_2012, Perez-Becker_2013, Booth_2023}. Intriguingly, the dust emission rates do not appear to follow an obvious pattern, and are instead hypothesized by \citet{Bromley_2023} to originate from chaotic mass-loss rates that are driven by surface vapor pressure, atmospheric hysteresis, and temperature feedback from the planetary atmosphere’s time-evolving optical depth.

Fig. \ref{fig:montage} also demonstrates clear variations in ingress and egress morphology across the individual transits of \system\,Ab. These variations are attributed to the angular distribution of dust grains relative to the planet's orbit as governed by the drift rate and the sublimation decay of grains over time. These aspects are explored in further detail in Section \ref{section:analytic}.

\subsection{(A)chromaticity of Fading Events}\label{section:achromaticity}
The multi-wavelength photometry collected in Section \ref{section:lco} enables a preliminary test of any distinct wavelength dependence of \system\,Ab's transits. Because the extinction efficiency of a dust grain diminishes as the grain circumference becomes smaller than the wavelength of incident light \citep{Hansen_1974}, a color dependence of transit depths may set constraints on dust grain sizes (e.g., \citealt{Croll_2014, Bochinski_2015, Sanchis-Ojeda_2015, Schlawin_2016, Colon_2018}). Chromaticity may also be expected in transit durations, in the sense that a trend of shorter durations towards redder wavelengths could suggest the sublimation of dust grains to smaller sizes further out in the tail.

\begin{figure}
    \centering
	\includegraphics[width=1.\columnwidth]{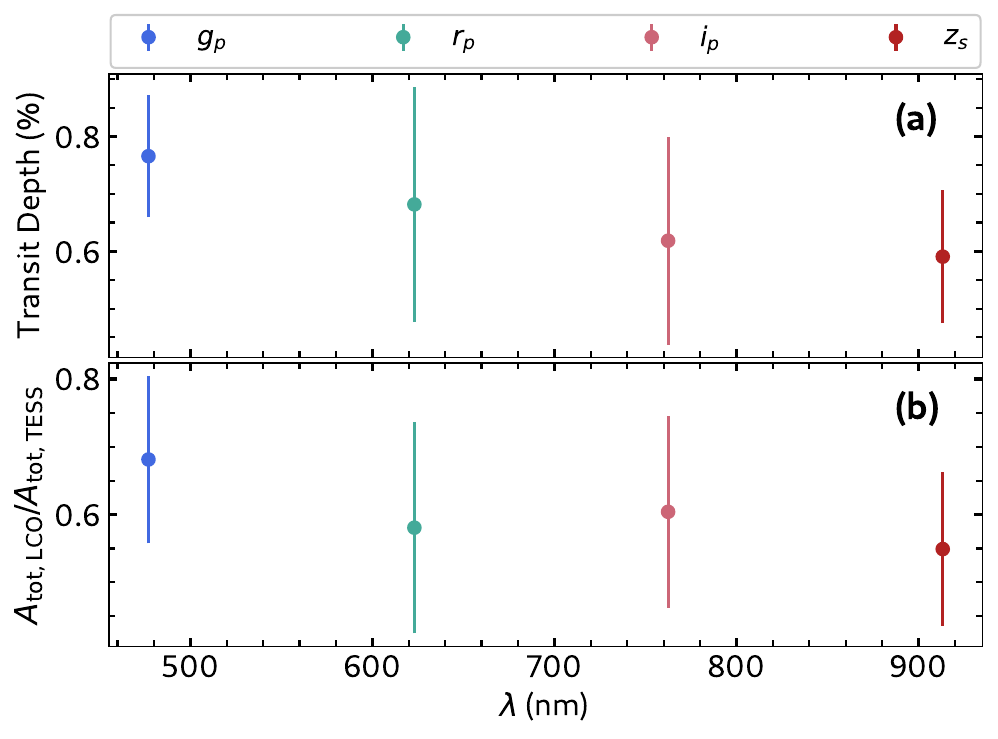}
    \caption{ Properties of the single transit observed with LCOGT multi-wavelength photometry. The points are plotted at approximately the central wavelength of the four MuSCAT3 bandpasses.  (a) A comparison of measured transit depths across the four LCO bandpasses. (b) Integrated area beneath each LCOGT light curve $(A_{\mathrm{tot, LCO}})$ relative to that measured from TESS $(A_{\mathrm{tot, TESS}})$. The total absorption is the integrated area beneath the model transit profile $\mathcal{T}(\phi)$ that is fit to the observed data.}
    \label{fig:lco_transit_areas}
\end{figure}

To test for transit chromaticity, we fitted $\mathcal{T}(\phi)$ to the LCOGT photometry across the four wavelength bands ($g_p, r_p, i_p, z_s$) individually and measured the transit depth of each light curve, as presented in Fig. \ref{fig:lco_transit_areas}a. In addition to transit depths, in Fig. \ref{fig:lco_transit_areas}b we show the integrated area beneath each fitted extinction profile ($A_{\mathrm{tot, LCO}}$) as a proxy for the total transit duration. By conducting two-sided t-tests, we found no statistically significant evidence for an increase in transit depths or an increase in $A_{\mathrm{tot, LCO}}$ towards shorter wavelengths. In particular, there is a 37\% chance that the observed trend in transit depths and a 50\% chance that the observed trends in $A_{\mathrm{tot, LCO}}$ could be due to random variation under the null hypothesis of no slope. These findings suggest a minimum grain radius of $\sim0.1$ microns, assuming a single grain size. \citet{Ridden-Harper_2018}, however, note that optically thick dust tails, resulting from high dust mass loss rates, may also produce achromatic transits. While we determine more robust constraints on grain sizes in Section \ref{section:grain_size}, we note that these results on the transits' wavelength dependence are preliminary, given only a single multi-wavelength series of observations to date. Further follow-up observations across a broader wavelength range and at a higher photometric precision will allow us to make more robust conclusions regarding the presence of any transit chromaticity.


\section{Dust Dynamics} \label{section:dust_dynamics}

\subsection{Analytic Calculations of Forward vs.~Trailing Dust Tails} \label{section:analytic}
The presence of forward and trailing dust tails for \system\,Ab provides unique constraints on $\beta$, the ratio of radiation pressure force to gravitational force for the dust grains. A dust grain that is released from the star-facing tip of the Hill sphere of the planet, $R_H$, with a velocity of $\Vec{v}$ will experience an effective gravitational force reduced by $(1-\beta)$, such that its specific energy is
\begin{equation} \label{eqn:ratio_a}
\begin{aligned}
    \mathcal{E}&=-\frac{GM(1-\beta)}{a_o-R_H} + \frac{1}{2}|\omega(a_o-R_H)\hat{V} + \Vec{v}|^2 \\
            & \equiv -\frac{GM(1-\beta)}{2a_d},
\end{aligned}    
\end{equation}
where $M$ is the mass of the host star and $a_d$ is the semi-major axis of the grain's new eccentric orbit, while $a_o$, $V$, and $\omega$ are the radius, orbital velocity, and angular frequency of the planet's orbit, respectively. A caveat with this formulation is that it ignores the small residual gravitational potential energy due to the planet. This residual energy is vanishingly small if the dust emerges from the vicinity of the planet's Hill sphere, unless, as explored in Section \ref{section:numerical}, the dust returns to the vicinity of the planet as their orbits cross. By adopting $GM/a_o = V^2$ and $\omega = V/a_o$, Eqn. \ref{eqn:ratio_a} can be rearranged to become:
\begin{equation}
\label{eqn:a_ratio}
    \frac{a_d}{a_o} = \frac{1-\beta}{2(1-\beta)/\chi - \chi^2 - 2\,\xi\,\chi(v/V)- (v/V)^2},
\end{equation}
where $\chi=(1-r_H)$ and $r_H\equiv R_H/a_o$ is the dimensionless Hill sphere radius. The parameter $\xi\equiv\sin\theta\cos\psi$ describes the angular dependence of the dust grain's initial vector with respect to the direction of the orbital motion $\hat{V}$. In particular, $\theta$ is the angle with respect to the line pointing from the planet to the host star, and $\psi$ is the azimuthal
angle around that vector, where $\psi=0$ aligns with the orbital velocity vector. Using Equation \ref{eqn:ratio_a}, the fractional difference in orbital period of the planet ($P_o$) and the grain ($P_d$) can be computed:
\begin{equation} \label{eqn:ratio_p}
\begin{aligned}
    \frac{P_d}{P_o} &= \bigg(\frac{a_d}{a_o}\bigg) ^{3/2}\frac{1}{\sqrt{1-\beta}}\\
    \Rightarrow \frac{\Delta P_d}{P_o} &= \frac{P_d}{P_o} - 1\\    
    &\simeq 2\beta + 3\,\xi(v/V) -6r_H,
\end{aligned}    
\end{equation}
where the approximation leverages $v/V\ll 1$, $\beta\ll 1$, and $r_h\ll 1$. Here, we apply the assumption that the drift rates of the grains are approximately constant and may be described by $\Delta P_d/P_o$ in Equation \ref{eqn:ratio_p}. Since values from the computed $\Delta P_d/P_o$ are generally small, the grains thus require numerous orbits around the host star to drift to angles as large as those observed. The angular drift of a dust grain from the planet (in units of $2\pi$) is $\Phi=N\Delta P_d/P_o$, where $N$ is the number of planetary orbits. Here, a positive value of $\Phi$ indicates a longer orbital period \textemdash{} material going into a trailing tail, and vice versa. 

We consider a scenario whereby a Parker wind deposits material that condenses into the dust grains at the Hill sphere radius $r_H = (M_p/3\,M)^{1/3}\simeq0.0032$, given a fiducial planet having mass $M_p=0.02\,M_\oplus$, radius 2,000 km, and escape velocity 3 km s$^{-1}$. We adopt the prescription from the \citet{Perez-Becker_2013} hydrodynamical wind models, whereby wind velocities approach the sound speed of the gas $c_s \simeq \sqrt{\gamma k\,T/\mu} $ at distances close to the Hill sphere radius. We use fiducial values of $\gamma=1.3$ and a mean molecular weight $\mu=30\,m_\mathrm{H}$ following the prescriptions by \citet{Perez-Becker_2013} and \citet{Booth_2023}, which are based on the atmospheric composition of hot, rocky planets reported from \citet{Schaefer_2009}. The sound speed at the surface of the planet is thus $c_s \sim 0.8\,$km s$^{-1}$, assuming a temperature of $T=1800\,\text{K}\sim T_{\mathrm{eq}}$. As shown by the \citet{Perez-Becker_2013} and \citet{Booth_2023} models, the outflowing gas expands and cools as it approaches the sonic point near the Hill sphere boundary. This leads to a reduction in its local sound speed relative to its value at the planetary surface, and implies a slower bulk flow speed at the sonic point. Consequently, the bulk velocity of the gas, $v$, is expected to be lower than $0.8\,$km s$^{-1}$ at the sonic point, and the models indicate that it may be only half this value. We therefore adopt $v=0.4\,$km s$^{-1}$ as the dust grain velocity at the Hill sphere.




Given the negligibly small magnitude of $v/V$ ($\sim$$0.4/180 \simeq 0.0022)$, Eqn. \ref{eqn:ratio_p}, which describes the drift of dust away from the planet, can be simplified to 
\begin{equation}\label{eqn:dust_length}
    \Phi\simeq N (2\beta - 6\,r_H),
\end{equation}
where as $\beta \to 0$, the maximum angular drift rate of grains in the leading tail, $\Phi_{\text{lead}}$, reaches a value of $6\,r_H$. Based on the modeling of the observed light curve in Section \ref{section:light_curve}, the observed ratio of tail lengths is known to be $\sim6$. We thus seek to find the maximum angular drift of grains in the trailing tail, $\Phi_{\text{trail}}$, which yields $\Phi_{\text{trail}}/\Phi_{\text{lead}}=6$. This corresponds to a maximum $\beta$ value (denoted $\beta_{\mathrm{max}}$), such that 
\begin{equation}
    \frac{\Phi_{\text{trail}}}{\Phi_{\text{lead}}} \simeq \frac{2\beta_{\mathrm{max}}-6\,r_H}{6\,r_H} = 6,
\end{equation}
which leads to $\beta_{\mathrm{max}}\simeq0.07$.

\begin{figure}
    \centering
	\includegraphics[width=1.\columnwidth]{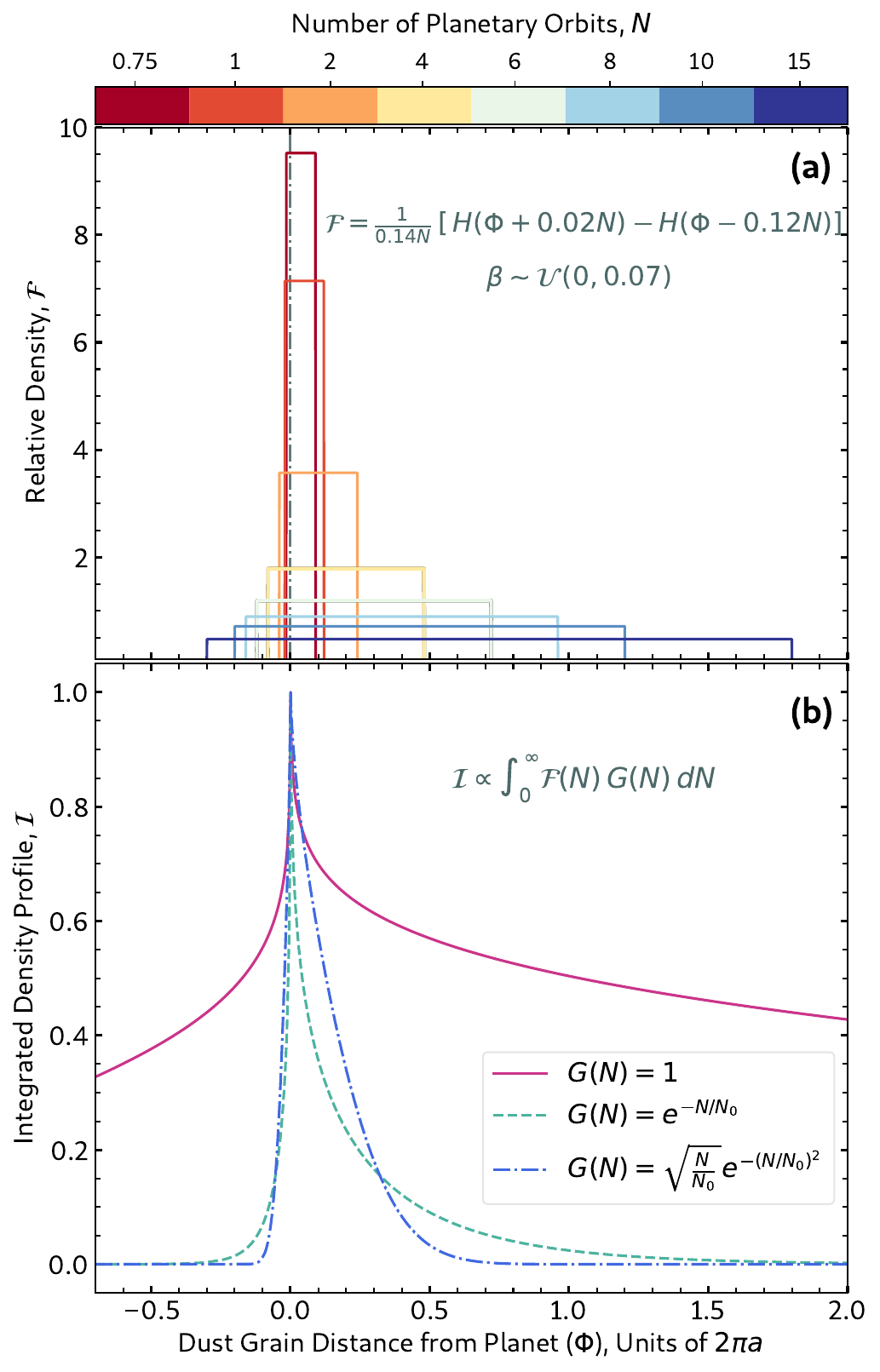}
    \caption{(a) Distribution of dust grain distances from the planet over time for an impulsive dust ejection event. The colored curves represent different numbers of elapsed orbits since the release. Values of $\beta$ are drawn from a uniform distribution $\mathcal{U}$, while $H$ denotes the Heaviside function. (b) A steady-state dust tail profile, shown as the integral of the dust distribution over time weighted by a decay function $G(N)$, characterized by a scale number $N_0$. The decaying profiles both use $N_0=2$, specifically chosen such that the resulting decay of the dust density visually resembles plausible transit profiles of \system\,A in Fig. \ref{fig:montage}. }
    \label{fig:analytic}
\end{figure}

Because $\Phi$ describes the angular spread of the dust grains, Fig. \ref{fig:analytic}a visualizes how an impulsive dust ejection event may evolve over a number of planetary orbits, assuming $\beta\sim\mathcal{U}(0, \beta_{\mathrm{max}})$. The distribution of dust in both trailing and leading dust tails is seen to slowly expand and drift over time.
Grains in the trailing tail, for instance, are predicted to spread by only $\Phi\sim0.1$ per orbit, which suggests that the orbit-to-orbit variations in the shape of the egress tails at $\Phi\sim0.2-0.5$ (see Figs. \ref{fig:montage} and \ref{fig:transit_stats}) are caused by attenuation from dust grains that were released 2-4 planetary orbits earlier, and have survived for that long. These dust grains, however, cannot survive indefinitely, because the observed transits experience an egress that transitions back to maximum flux in the light curve. As dust grains drift from the planet, they are expected to eventually sublimate at rates that depend on their sizes, compositions, and temperatures (see, e.g., \citealt{vanLieshout_2014}). Consequently, steady-state analytical profiles of the dust tail require weighting by a decay function to allow grains to leave the system. In Fig. \ref{fig:analytic}b, we show qualitatively that decay functions in the form of exponential or Gaussian-like functions may yield dust density profiles whose decay with angular distance from the planet resembles those seen in the observed data. These functions are only illustrative, given the currently unknown dust grain size distribution and grain composition.

\subsection{Numerical Simulations of Dust Tails} \label{section:numerical}

\begin{figure*}
    \centering
	\includegraphics[width=2.\columnwidth]{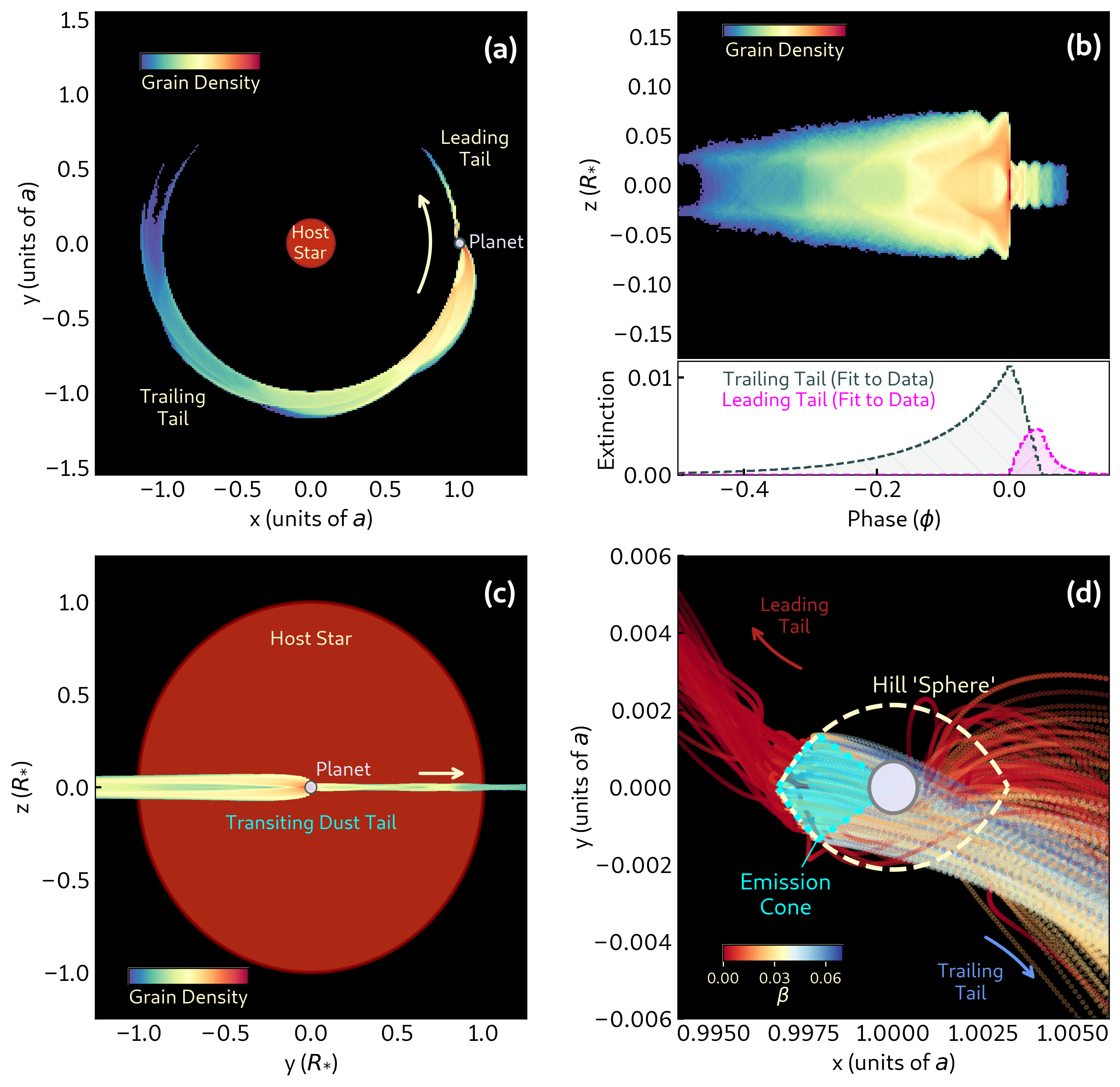}
    \caption{3D simulation of 50,000 dust grains in the co-rotating frame of the orbiting planet, integrated for up to 30 planetary orbits. The host star is shown to scale, while the planet's size is inflated by a factor of 50 in panels (a) and (c) for visual clarity. (a) A view from above the planetary orbit, looking down at the $x-y$ plane in which the planet is orbiting counterclockwise. The trails indicate the accumulated trajectories of the dust grains over time. There are two distinct trails that correspond to the leading and trailing dust tails. (b) A view of the accumulated trajectories of the dust tails in the $z-\phi$ plane (an `unrolled' dust tail), showing the azimuthal extent of the tails. For comparison, the subpanel beneath the color image shows the absorption profiles of the leading and trailing tails, which are based on fitting the mean transit profile $\mathcal{T}(\phi)$ to the TESS data in Fig. \ref{fig:transit_model}. (c) Visualization of the simulated dust tails projected onto the $z-y$ plane, with the host star shown in the background to demonstrate the extent of the dust tail's height. (d) A close-in view of the dust tail in the $x-y$ plane. Tracks correspond to trajectories of individual grains as they are launched from the boundary of the Hill `sphere' of the planet. The region in cyan represents the outflowing Parker wind. Each grain is assigned a $\beta$ value, which influences its trajectory and whether it ends up in the leading or the trailing tail. }
    \label{fig:tail_sim}
\end{figure*}

To visualize the dust tails, we performed a series of numerical simulations following the approaches presented by \citet{Rappaport_2012} and \citet{Sanchis-Ojeda_2015}. In particular, we simulated the release of 50,000 dust grains and numerically integrated their trajectories in the co-rotating frame of the orbiting planet. We assumed a nominal planet with a mass of 0.02 M$_\oplus$ and a radius of $2000\,$km that is orbiting a $0.7M_{\odot}$ host star at a scaled semi-major axis of $a/R_{*}$ of 6.4.

Grains were released in random directions within a 1 $\, \text{sr}$ emission cone centered on the direction of the host star. Similar to Section \ref{section:analytic}, we adopted an initial velocity of $v = 0.4\,$km s$^{-1}$ at the planet's Hill radius, as appropriate for a thermal Parker hydrodynamically driven outflow \citep{Rappaport_2012, Perez-Becker_2013}. Instead of a spherical boundary from which the grains are released, we adopted the actual shape of the Hill `sphere' following \citet[their Appendix A]{Rappaport_2016}, which describes the critical potential surface, which has a biconvex shape around the tidal axis ($\hat{x}$). Upon launch, each grain's trajectory is integrated for 30 planetary orbits in 6000 steps per orbit using a 4$^{\text{th}}$ order Runge-Kutta integrator. During the simulation, any grain whose trajectory falls back onto the planetary surface is removed as it will not contribute further to the dust tails.

\subsubsection{Grain Motion in the Orbital Plane}

An illustrative set of results is shown in Fig.~\ref{fig:tail_sim}.  Panel (a) shows a view looking down on the orbital ($x-y$) plane where it is evident that both leading and trailing dust tails readily form from the released grains. To match the angular extent of both tails from the observed data \textemdash{} approximately $180^{\circ}$ in orbital phase from ingress to egress \textemdash{} we assume all grains have extinction cross-sections with a fixed exponential decay rate comprising an $e$-folding time of 2 planetary orbits. A critical factor of these simulations is the assigned value of $\beta$ for each grain, drawn from a uniform distribution whereby $\beta\sim\mathcal{U}(0, 0.07)$.  This is because the leading tail is only populated by grains with $\beta\apprle0.01$, consistent with the values that will yield $\Phi<0$ in Eqn. \ref{eqn:dust_length}. All grains with small $\beta$ are initially released into smaller and faster orbits that will {\it lead} the planet in their motion---and they will remain in those orbits. Conversely, grains that experience higher levels of radiation pressure (higher $\beta \simeq 0.01-0.07$) experience a `backflow' after release and are pushed into higher and slower orbits that will lag the planet --- thereby forming the trailing tail.

Not only is the leading tail shorter than the trailing tail, but it is also more narrow in the radial direction. The widths of the tails are governed by the range of semi-major axes of the dust grains and their orbital eccentricities.  The latter can be calculated analytically, but the expression is not particularly amenable to expansion for small $\beta$, $v/V$, and $r_H$.  However, by evaluating the exact expression numerically and averaging over the emission directions (as described by the parameter $\xi$ in equation \ref{eqn:a_ratio}), we find that the eccentricity of the dust is given approximately by
\begin{eqnarray}
\label{eqn:eccewntricity}
e_d & \,\approx & \,0.01 -0.86 \, \beta  ~~ {\rm for} ~~ \beta \lesssim 0.01 \nonumber \\
e_d & \,\approx & \,-0.01 + 1.04 \, \beta  ~~ {\rm for} ~~ \beta \gtrsim 0.01.
\end{eqnarray}
Because the dust grains in the leading tail have $\beta \lesssim 0.01$, the eccentricities in the leading tail range from $0.002 \lesssim e_d \lesssim 0.01$. In comparison, grains in the trailing tail have eccentricities between $0.01 \lesssim e_d \lesssim 0.06$.  

\begin{figure}
    \centering
	\includegraphics[width=1.\columnwidth]{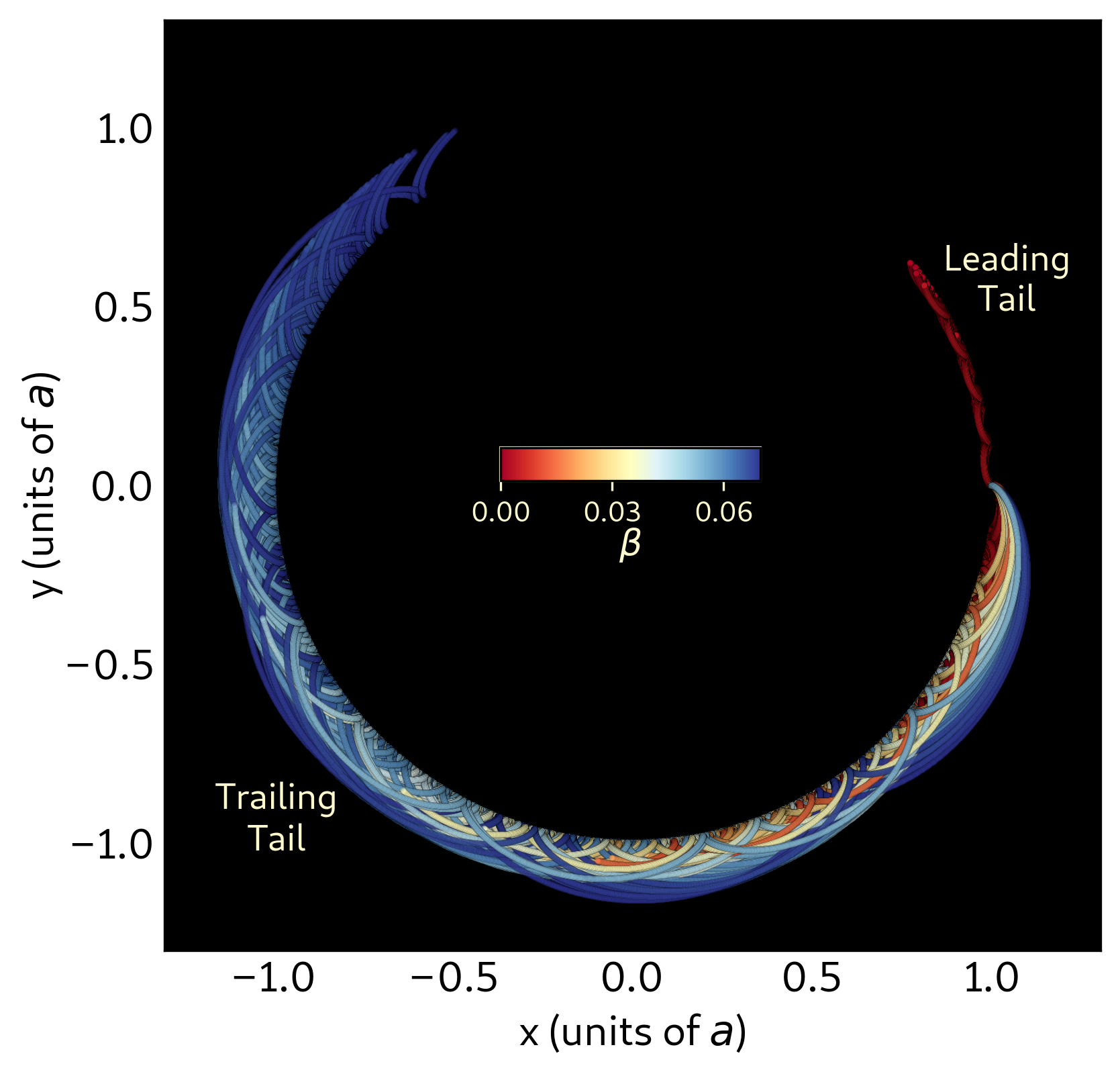}
    \caption{ Trajectories of simulated dust grains as viewed in the $x-y$ plane, with each trajectory colored by the $\beta$ value of its corresponding grain. The parameters of this simulation are similar to those presented in Fig. \ref{fig:tail_sim}, except that the trajectories here are numerically integrated across only 5 orbits, and the grains are not allowed to decay.}
    \label{fig:tail_widths}
\end{figure}

Altogether, the range of dust orbital distances from the host star (in units of $a_o$) will have two contributions: (i) the range of distances due to the eccentricity ($\simeq 2 e_d$), and (ii) the range of semi-major axes.  For the latter, equation (\ref{eqn:a_ratio}) yields 
\begin{equation}
\frac{a_d}{a_o} \simeq 1+\beta-4\,r_H+2\xi v/V,
\end{equation}
from which we find that the fractional variations in semi-major axes are given just by $\beta$ if the $v/V$ term is neglected.
Therefore, the combined variations in the radial width $\Delta r$ are given roughly by
\begin{equation}
\frac{\Delta r}{a_o} \approx \beta + 2e_d \approx 3e_d.
\end{equation}
Grains in the trailing tail with larger $e_d$ will thus have $\Delta r$ of $\sim$10\%-15\%, as opposed to grains in the trailing with $\Delta r$ of only a few percent. This result is illustrated in Fig.~\ref{fig:tail_widths}, where it is evident that high-$\beta$ grains (in blue) contribute the most to $\Delta r$ of the radially-wide trailing tail, whereas grains with the smallest $\beta$ (in red) form the radially-thin leading tail.

\subsubsection{Dust Tail Heights}
The view of the simulated dust tail as it passes by the host star is shown across different scales in panels (b) and (c) of Fig.~\ref{fig:tail_sim}.  The vertical extent of the trailing tail is about $\pm 0.05\,R_{\star}$, while that of the leading tail is about a factor of three times smaller.  A key question is: what determines the vertical height? Using the same approximations and assumptions used to derive equations (\ref{eqn:ratio_a}) and (\ref{eqn:ratio_p}), the tilt $\zeta$, of an orbit with respect to the planet's orbital plane is given by 
\begin{equation}
\zeta \simeq (v/V) \sin \theta \sin \psi < (v/V)
\end{equation} 
Given that the orbital speed of the planet is $\simeq 175$ km s$^{-1}$ and that of the Parker-wind outflow is $v=0.4$ km s$^{-1}$, the maximum tilt is $\zeta \lesssim 0.002$ rad.  Meanwhile, the projected disk height in units of $R_{\star}$ is given by 
\begin{equation}\label{eqn:disk_height}
\frac{h}{R_*} \simeq (v/V) (a/R_*) \simeq 0.014
\end{equation}
This is consistent with the heights observed in the leading tails (Fig. \ref{fig:tail_sim}b) and is barely adequate to explain the typical 1\% deep transits that are observed.  However, the trailing tail is more than sufficiently high to account for the observed transits.  Fig. \ref{fig:tail_sim}c shows the same view of the tail crossing the stellar disk, but in distance units that are identical across the $y$ and $z$ directions.  This view demonstrates the narrowness of the dust tail's vertical extent.

Of particular interest is how the simulated trailing tail may gain a vertical extent up to $\sim0.05R_{\star}$, given that $h\simeq0.014\,R_{\star}$. 
Fig. \ref{fig:tail_sim}d examines the grains' trajectories in the vicinity of the planet, where the Hill ``sphere'' is shown to scale, with its proper shape.  The emission cone that we assumed for the Parker wind is shaded in cyan, and the grains are all released on the Hill surface with speed $v$ outward.  The grains with small $\beta$ glide into the leading tail (toward the top of the figure). In contrast, grains with $0.01 \lesssim \beta \lesssim 0.07$ experiences a `backflow' as described previously, which causes the grains to drift back though the Parker wind outflow region \textemdash{} an outcome we view as problematic for this particular launch scenario.  
Among these backflowing grains, those that initially have significant altitudes above the orbital plane are scattered by the planet, such that they are boosted (i.e., gravity assisted) to larger heights below the planet's orbital plane than would otherwise be attained \textemdash{} and vice versa. While this explains how the large (and necessary) heights are obtained in this simulation, it is challenging to accept because of the apparent conflict between the outflowing Parker wind and the back-tracking grains that form the trailing tail.

In Appendix \ref{appendix:alternative_launches}, we explore varying the grain turn-on time in our simulation. We find a scenario in which dust grains form half an orbit after launch \textemdash{} simulating a later decoupling of the dust grains from the outflowing gas \textemdash{} yields smaller tail heights (up to $\sim0.02\,R_{\star}$) and a leading tail with the same vertical height as the trailing tail. Such a scenario would require optically thick ($\tau\sim2$) dust grains to reproduce the observed transit properties of \system\,Ab and would not allow for transits deeper than $\sim2\%$ should they be observed in the future. 

We note that the conclusions from our simulations depend only marginally on our assumption of $v= 0.4$ km s$^{-1}$ as the launch velocity of the dust grains. For higher values of $v$, as shown for some specific models by \citet{Perez-Becker_2013} and \citet{Booth_2023} under certain conditions, the most salient outcome of this change would be that the value of $h/R_{\star}$ from Eqn. \ref{eqn:disk_height} would double to 0.028. In this case, the tail heights inferred from this analytic description are now largely adequate to explain the observed transit depths above $1\%$. To observe the effect of a higher launch velocity on our simulations, we considered another alternate scenario in Appendix \ref{appendix:alternative_launches}, whereby we assumed the dust grain release to be from the planetary surface with the surface escape velocity ($v=3\,$km s$^{-1}$) \textemdash{} similar to a scenario explored by \citet{Ridden-Harper_2018}. As expected, these conditions may form the requisite dust tails, but the nature of the launch is incommensurate with a Parker wind scenario. Our simulations, while mainly illustrative, indicate that accurately reproducing the observed transit profiles will require not only the detailed modeling of outflow dynamics and dust formation processes \citep{Perez-Becker_2013, Croll_2014, Booth_2023} but also a detailed consideration of the post-release behavior of dust grains \citep{Ridden-Harper_2018} and variations in optical depth along the dust tail \citep{Campos_Estrada_2024}.

\begin{figure}
    \centering
	\includegraphics[width=1.\columnwidth]{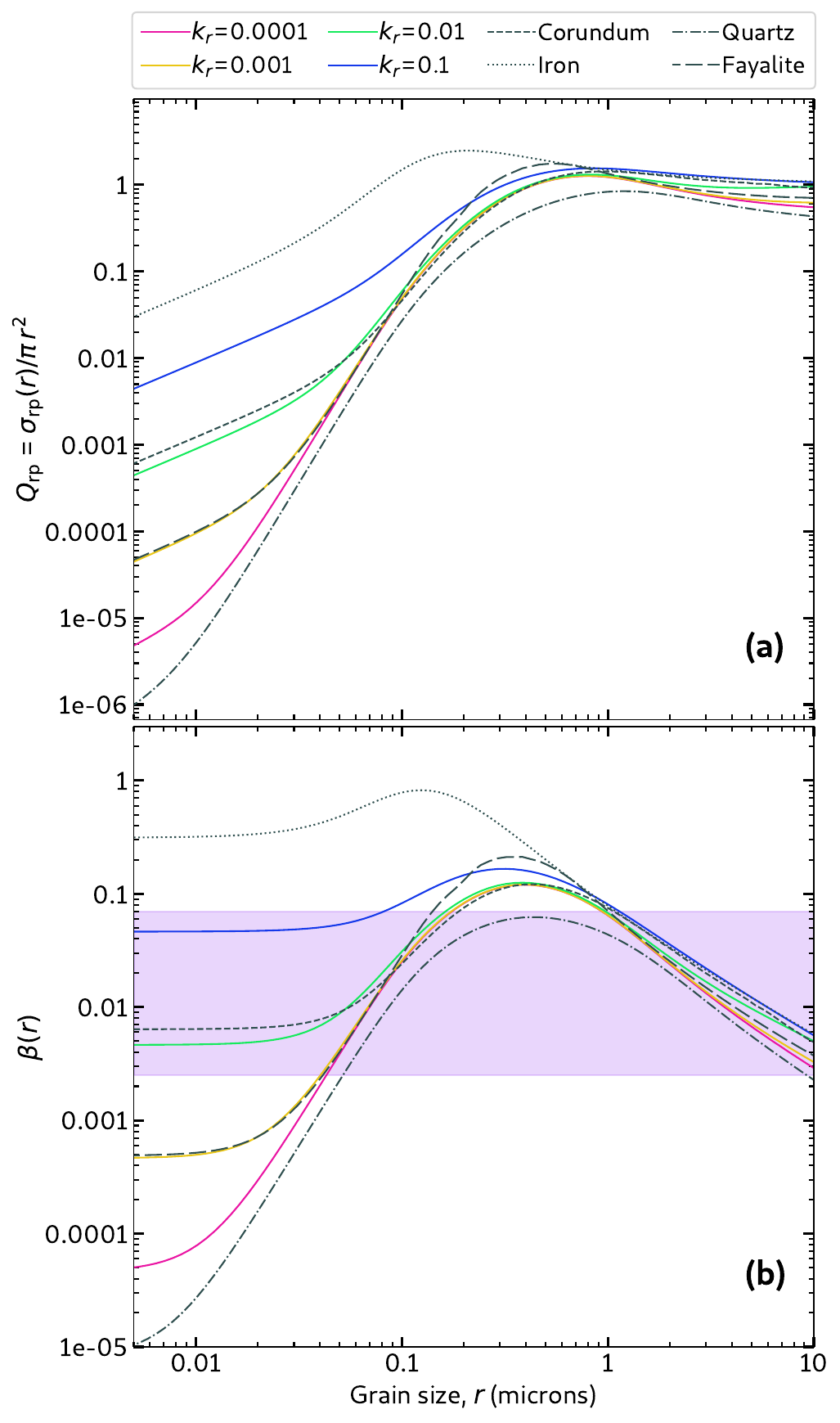}
    \caption{(a) Radiation pressure cross-sections calculated using Mie theory. The cross-sections are normalized by the geometric cross section of the dust grain and are displayed as a function of the grain size (radius) $r$. Solid curves show cross-sections for illustrative materials with a real part of the index of refraction $n_r=1.65$ across a range of $k_r$, the imaginary part of the index of refraction. Cross-sections for representative astronomical minerals are displayed as non-solid curves. (b) Calculated values of $\beta$ from Eqn. \ref{eqn:beta}, as a function of grain size for the same materials presented in (a). The purple band corresponds to a plausible range of $0.0025<\beta<0.07$ for grains in the trailing tail, as inferred from \system\,Ab's transits.}
    \label{fig:materials}
\end{figure}

\subsection{Inferences on Grain Size}\label{section:grain_size}
From our inferred distribution of $\beta$, we may constrain the sizes of the dust grains. This follows from the definition of $\beta$:
\begin{equation}\label{eqn:beta}
    \begin{aligned}
    \beta\equiv\frac{F_{\mathrm{rad}}}{F_{\mathrm{grav}}}&=\bigg[\frac{\sigma_{\mathrm{rp}}(r,\lambda)L}{c\,4\pi d^2}\bigg]\bigg[\frac{GmM}{d^2}\bigg]^{-1} \\
    &= \frac{3 Q_{\mathrm{rp}}(r, \lambda)L}{c\,16\pi\rho r  GM},
    \end{aligned}
\end{equation}
where $\sigma_{\mathrm{rp}}(r,\lambda)$ is the radiation pressure cross-section for a dust grain (assumed to be spherical) of size $r$  at wavelength $\lambda$, $Q_{\mathrm{rp}}(r,\lambda)$ is the dimensionless form of the cross-section in units of $\pi r^2$, and $\rho = 3\,\text{g}\,\text{cm}^{-3}$ is the assumed bulk density of the grain material. To calculate $\sigma_{\mathrm{rp}}(r,\lambda)$, we used the \texttt{miepython}\footnote{\url{https://miepython.readthedocs.io}} package, which follows the Mie scattering procedures outlined by \citet{Wiscombe_1979}.
To determine $\sigma_{\mathrm{rp}}(r)$, namely cross-sections dependent only on $r$, we integrated $\sigma_{\mathrm{rp}}(r,\lambda)$ over a Planck curve evaluated at the host star's $T_{\mathrm{eff}}$ as an approximation to the star's spectral energy distribution. These results are shown in Fig. \ref{fig:materials}a for several materials representative of minerals within a planetary crust and mantle (e.g., \citealt{Hirose_2013, Mahapatra_2017}). We accessed the real ($n_r$) and imaginary ($k_r$) parts of the indices of refraction for iron \citep{Johnson_1974}, quartz \citep[fused silica, SiO$_2$]{Franta_2016}, and corundum \citep{Querry_1985} from the \href{https://refractiveindex.info/}{\texttt{RefractiveIndex.INFO}} repository \citep{Polyanskiy_2024}. For fayalite (Fe$_2$SiO$_4$), we retrieved the corresponding optical constants from \citet{Fabian_2001}. For comparison, we also show the curves for generic materials with $n_r=1.65$ and $k_r\in[0.0001, 0.001, 0.01, 0.1]$, which are assumed to be independent of $\lambda$.

Fig. \ref{fig:materials}b shows $\beta$ as a function of grain size for \system\,A. We find that $0.0025<\beta<0.07$ requires either large grain sizes ($ 1\,\mu{\rm m} \lesssim r  \lesssim 10\,\mu$m) for essentially any grain composition, or small grain sizes ($\lesssim 0.05 \, \mu$m) in tandem with very specific mineral compositions. Small grains tend to readily absorb the short wavelengths of the host star but radiate weakly in the infrared.  As a result, they sublimate much quicker than larger grains (see, e.g., \citealt{Xu_2018}), which suggests that the dust grains emanating from \system\,Ab are predominantly large.



\subsection{Mass Loss Rate}
We can provide rough estimates of the mass loss rate in dust from the planet, given the inferred range of $\beta$ and measurements of the dust tail's length. We first determine the dust lifetime, $T_d$, by estimating the characteristic length for the observed trailing tail (distance to its $1/e$ point) to be $\phi_c\sim0.1$, or about 10\% of the circumference of the planetary orbit (Appendix \ref{appendix:individual_transits}).
The average drift of dust grains in the trailing tail is $\langle\Phi_{\mathrm{trail}}\rangle \simeq (2\langle\beta\rangle- 6r_H)N$, where, again, $N$ is the number of orbits since launch, $\langle \beta \rangle \simeq 0.03$, and $6r_H \simeq 0.02$.  Thus, $\langle \Phi_{\rm trail}\rangle \simeq 0.04 N$. Finally, if we equate $0.04 N$ to 0.1 (orbital circumferences), we find a mean lifetime for the dust grains of $\sim$2.5 orbits, or about 3 days.

The mass of dust at a given instant in the trailing tail can be estimated by `unrolling' the tail into a sheet of dust of height $\mathcal{H}$, length $\mathcal{L}$, and depth $\mathcal{D}$. In the optically thick limit, $\mathcal{H}=2\cdot\langle\delta\rangle\cdot R_{\star}$, where $\langle\delta\rangle=0.012$ is the average transit depth. Meanwhile, $\mathcal{L} = \phi_c\cdot 2\pi R_{\star}\cdot (a/R_{\star})$, and $\mathcal{D}\approx 2r=10\,\mu$m, where $r \approx 5\,\mu$m is an adopted average grain radius. Assuming $\rho=3\,$g\,cm$^{-3}$ as the bulk density of the grain, we find that the approximate mass in the dust ribbon is

\begin{equation}
    \begin{aligned}
        M_{\mathrm{dust}} &\approx \rho\mathcal{H}\mathcal{L}\mathcal{D}\\
                          &\approx 6\times10^{17}\left(\frac{s}{5\,\mu{\rm m}}\right)\left(\frac{\rho}{3\,{\rm g\,cm}^3}\right) \left(\frac{R_{\star}}{0.7 \, R_\odot}\right)~~{\rm grams}
    \end{aligned}
\end{equation}

As the sublimated grains in the tail are steadily replenished by mass loss from the planet, the rate of mass loss from the planet is calculated as:
\begin{equation}
    \begin{aligned}
        \dot{M}_{\mathrm{dust}} &\approx  M_{\mathrm{dust}}\cdot T_d^{-1} \\
                &\approx 2 \times 10^{12} ~{\rm g\,s^{-1}} \simeq 10 \, M_{\rm \oplus} \,{\rm Gyr}^{-1}.
    \end{aligned}
\end{equation}
Assuming a planet radius of 2,000 km and that the sub-stellar region covers a surface area corresponding to a solid angle of 1 $\, \text{sr}$, $\dot{M}_{\mathrm{dust}}$ corresponds to a mass loss flux, $J \simeq 5 \times 10^{-5}\,{\rm g\,cm^{-2}\,s^{-1}}$. To identify if a planetary surface may vaporize at such a rate, we approximate values of $J$ as a function of temperature in a vacuum following the formulation by \citet{vanLieshout_2014}. In particular, it assumes an equilibrium between the rates of material entering and leaving the vapor from a solid surface \citep{Langmuir_1913}, and represents the limiting case in which sublimated matter is swept away from the planetary surface as quickly as it is emitted. By adopting the sublimation parameters in Table 3 of \citet{vanLieshout_2014}, we find that  iron, silicon monoxide (SiO), and crystal fayalite (Fe$_2$SiO$_4$) have $J\sim2-15 \times 10^{-5}\,{\rm g\,cm^{-2}\,s^{-1}}$ at $T_{\mathrm{eq}}$, consistent with the mass loss flux for \system\,Ab.

To estimate a remaining lifetime for the planet, we again assume a fiducial planet mass of $0.02\,M_{\oplus}$. This choice of planetary mass is reasonably well-justified, given that \citet{Perez-Becker_2013} demonstrate that planets with high $\dot{M}$ are generally those in the low-mass regime ($M_p \apprle 0.02\,M_{\oplus}$) in their final, short-lived evaporating phase. Based on our estimates of $\dot{M}$, we thus infer a remaining lifetime of $0.002$ Gyr $=2$ Myr for \system\,Ab, indicating only a brief duration before the planet's complete disintegration.

\section{Discussion and Outlook}
\system\,Ab presents an extreme case of a catastrophically evaporating rocky planet. Compared to the three other known disintegrating planets, \system\,Ab has the deepest transit depths and longest tails, indicative of a rapidly vaporizing planet emanating copious amounts of dust. Of particular interest is its long orbital period and low equilibrium temperature, $T_{\mathrm{eq}} = 1820\pm 45\,$K relative to the other disintegrating planets ($T_{\mathrm{eq}}\sim2100\,$K), which may be suggestive of dust properties (e.g., composition, sublimation conditions) that are distinct from other such systems.

Unique also to \system\,Ab is the clear presence of extended trailing and leading tails, each contributing to the net absorption of starlight from the host star. It is this property that has enabled us to develop a framework for estimating the range of $\beta$ (i.e., up to 0.07) \textemdash{} and subsequently to infer large grain sizes of $\approx 1-10\,$microns \textemdash{} based on the drift of dust grains along both tails. As demonstrated by our simulations in Section \ref{section:numerical}, the added presence of the leading tail can also be particularly valuable for constraining the dynamics of the dust tail and for testing dust formation scenarios. 

Notable in \system\,A's light curve, as well, is the absence of a forward scattering peak, a feature seen in the transits of the other disintegrating planets \textit{Kepler}-1520~b and \textit{K2}-22~b. 
It is likely that the flux enhancement from forward scattering is canceled out by the light extinction of the transit. More specifically, micron-sized grains are mainly scattered within a narrow cone in the forward direction to an observer, thus remaining within highly absorbing regions of the transit. In addition, \citet{Devore_2016} estimated that the contrast between the scattering amplitude and transit depth scales with $\pi (R_{\star}/a)^2$. Any scattering feature within \system\,A's light curve would be confined within the deepest parts of the transit, with an amplitude that is about a factor of 15 smaller than the depth of the transit. 

\begin{figure}
    \centering
	\includegraphics[width=1.\columnwidth]{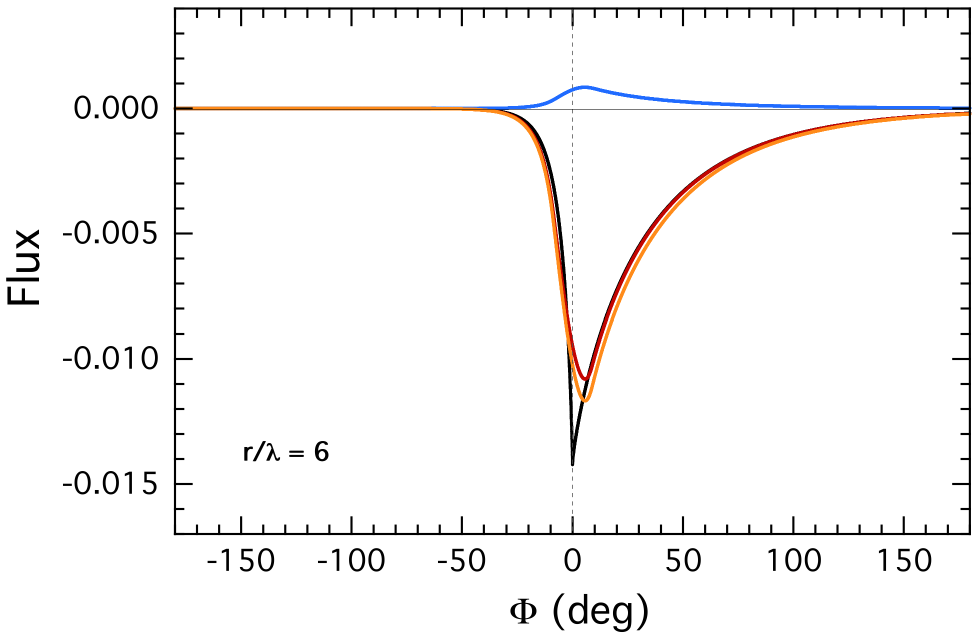}
    \caption{Simulated contributions to the transit profile. The black, orange, blue, and red curves are the dust extinction profile, the absorption profile of the transit, the forward scattering peak, and the net observed transit profile. These results were computed using a representative value of $r/\lambda = 6$, and are generally insensitive to varying $r/\lambda$ between 1-10.}
    \label{fig:forward_scatt}
\end{figure}

We show an illustrative example in Fig.~\ref{fig:forward_scatt} of the possible effect of forward scattering on the transit light curve. The black curve represents what the dust extinction profile might look like during a typical transit.  The orange curve is the corresponding observed absorption profile after the dust tail has been convolved with the limb-darkened stellar disk. The blue curve represents the forward scattering contribution under the assumption that the dust optical thickness is $\lesssim 0.5$ everywhere in the tails. A larger optical thickness would only diminish the forward scattering peak further. The net observed transit after the forward scattering is added to the absorption profile is shown by the red curve and demonstrates that discerning the presence of any forward scattering within the observed transit would be difficult.


Based on the models by \citet{Perez-Becker_2013}, it is plausible that \system\,Ab began with an initial mass greater than Mercury's, but experienced evaporative mass-loss spanning several Gyr. The lack of observational evidence for strong stellar activity suggests a quiescent host star, which when coupled with its derived isochronal age in Table \ref{table:properties}, suggest a planetary system older than the Sun. While the formation scenarios for disintegrating rocky planets may be similar to those proposed for USPs (\citealt{Winn_2018}, and references therein), one possibility, as suggested by \citet{Sanchis-Ojeda_2015}, is the role of a binary companion in driving the planet toward the host star via von Zeipel–Lidov-Kozai cycles with tidal friction \citep{vonZeipel_1910, Lidov_1962, Kozai_1962}. Notably, \system\,Ab is the second disintegrating planet whose host is part of a known wide binary system. The other is the \textit{K2}-22 system, in which the host star primary orbits another M-dwarf at a projected physical separation of $\sim430$ AU. Another possibility is the presence of star-planet interactions in the form of magnetic drag from the motion of the planet through the stellar magnetosphere. As described by \citet{Lee_2025}, such an interaction induces a torque that can efficiently decay the orbits of small, rocky planets from farther out, bringing them within range of disintegration. The discovery of more disintegrating rocky planets with TESS and potentially from PLATO \citep{Rauer_2014} will certainly be helpful to pin down the pathways leading to the demise of these planets.

\system\,Ab presents a compelling target for exoplanet mineralogy studies because its dusty effluents form from the mineral vapor that sublimated off the planetary surface. In support of this, \citet{Curry_2025} demonstrated that the mineral composition of the dust tails may directly reflect those from the mantle of the planet that have melted in lava pools on its surface, based on simple models of lava pool-atmosphere systems.
Transmission spectroscopy of the dusty effluents therefore provides an opportunity to directly measure the surface composition of a small, rocky exoplanet. Considerable effort has been expended for ground-based follow-up observations to probe the mineral composition of the dust (or gas) emanating from two of the three previously known disintegrating planets (e.g., \citealt{Croll_2014, Bochinski_2015, Colon_2018, Gaidos_2019, Ridden-Harper_2019}). Such observations, however, have been challenging due to the faintness of those host stars in tandem with their small transit depths ($\sim0.5\%$). At optical wavelengths, \system\,A is $\sim$$100$ times brighter than \textit{K2}-22, and is $\sim$$250$ times brighter than \textit{Kepler}-1520. The relatively deep transits of this system, which have an average depth of $\sim1.2\%$, are expected to provide high signal-to-noise measurements from existing ground-based facilities. Furthermore, the weak, yet variable transits of previously known disintegrating planets sometimes produce observational intervals with no detectable transits \citep{vanWerkhoven_2014, vanLieshout_2018}. This is in contrast to the consistently deep and long transits shown by \system\,Ab, which greatly increases their detectability as evidenced by our follow-up measurements in this work (Fig. \ref{fig:lco}).

Beyond ground-based efforts, observations of \system\,Ab from the James Webb Space Telescope (JWST) may reveal exceptional new insights on planetary interiors. Low-resolution spectra from JWST's Mid-Infrared Instrument (MIRI) are well-suited for detecting resonant features expected from minerals comprising terrestrial planets. A transit observation having a $\sim1\%$ depth even from targets as faint as \textit{K2}-22 may reveal the presence of mineral species characteristic of specific interior layers of terrestrial planets \citep{Bodman_2018, Okuya_2020}. Pioneering results by \citet{Tusay_2025} demonstrated this capability observationally by reporting detected mineral and gaseous features in \textit{K2}-22's mid-infrared JWST transmission spectra even from transits as shallow as 1,600 ppm. With \system\,A being $\sim100$ times brighter than \textit{K2}-22, the potential for composition measurements is enormously higher and may enable even more refined geological differentiation (e.g., oceanic versus continental crust) based on distinct elemental ratios measured from minerals detectable in the mid-infrared like Fe, Mg, and Si (e.g., \citealt{Rudnick_2003}).

Finally, we note that the relative brightness of \system\,Ab opens up possibilities for polarization measurements of light scattered from its dust tails, which may provide additional or complementary information to photometry and spectroscopy regarding the structure of the dust tail (e.g., \citealt{Stinson_2016}) as well as the size and optical properties of its constituent dust grains (e.g., \citealt{Zubko_2015}). The detection of polarization would further demonstrate the existence of the forward scattering feature in the dust tail, which is currently obscured within the larger absorption regions of the transit.

\appendix

\section{Mass Limits of Detectable Companions from NEID RVs} \label{appendix:rv_lim}
\begin{figure}
    \centering
	\includegraphics[width=0.5\columnwidth]{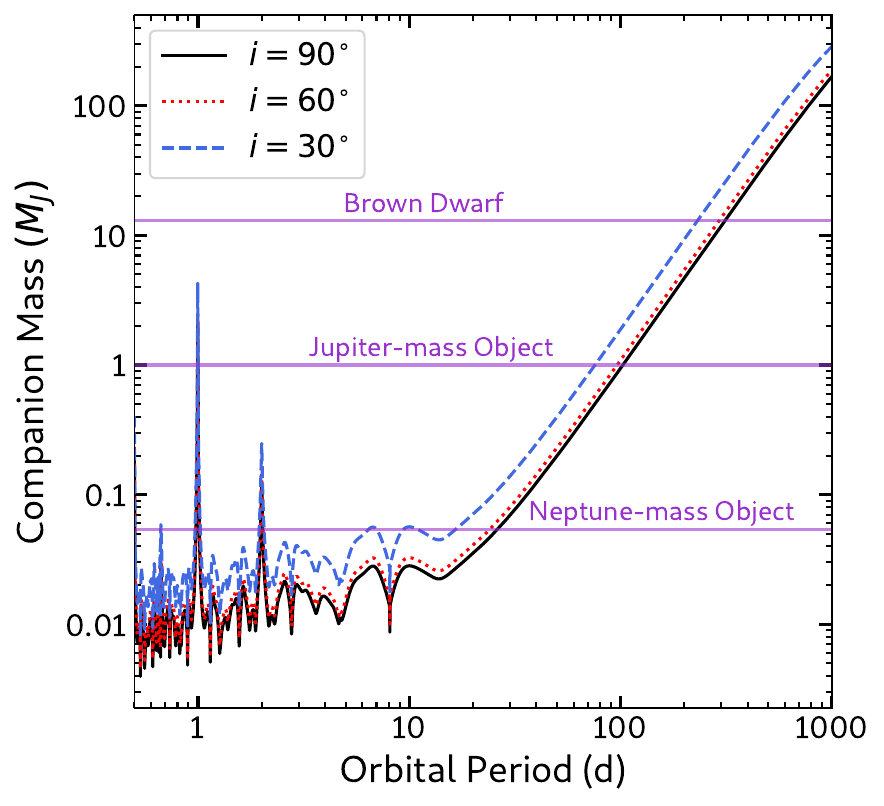}
    \caption{Detection limits of the NEID radial velocity (RV) observations in Section \ref{section:neid}, below which the RV data is insensitive to the presence of a companion to \system\,A. Shown are curves for three orbital inclination angles ($i$) of the companion. The spikes at 1 and 2 days are caused by the RV observations being taken at increments of a day.}
    \label{fig:rv_lims}
\end{figure}

To determine the upper allowed mass limits for companions orbiting from the seven NEID radial velocity (RV) observations, we performed least-squares fitting of assumed circular orbits to the observations across 400,000 trial periods between 0.5 and 1000 days. Each orbit contains three free parameters (RV semi-amplitude $K_1$, orbital phase, and RV offset). At each trial period, the best-fitting $K_1$ is incremented by its $1$-$\sigma$ uncertainty, then used to compute an upper limit to the mass function $f(M)$ at that period. By considering orbital inclinations of $i=30^{\circ},60^{\circ}, 90^{\circ}$, the function $f(M)$ is converted into limits on the mass of a hypothetical planet by solving for $M_{\mathrm{comp}}$ in the equation $f(M) = (M_{\mathrm{comp}}\sin i)^3/(M_{\star}+M_{\mathrm{comp}})^2$, as presented in Fig. \ref{fig:rv_lims}.

\section{Fits to Mean Transit Profile} \label{appendix:posterior}
\begin{figure}
    \centering
	\includegraphics[width=0.75\columnwidth]{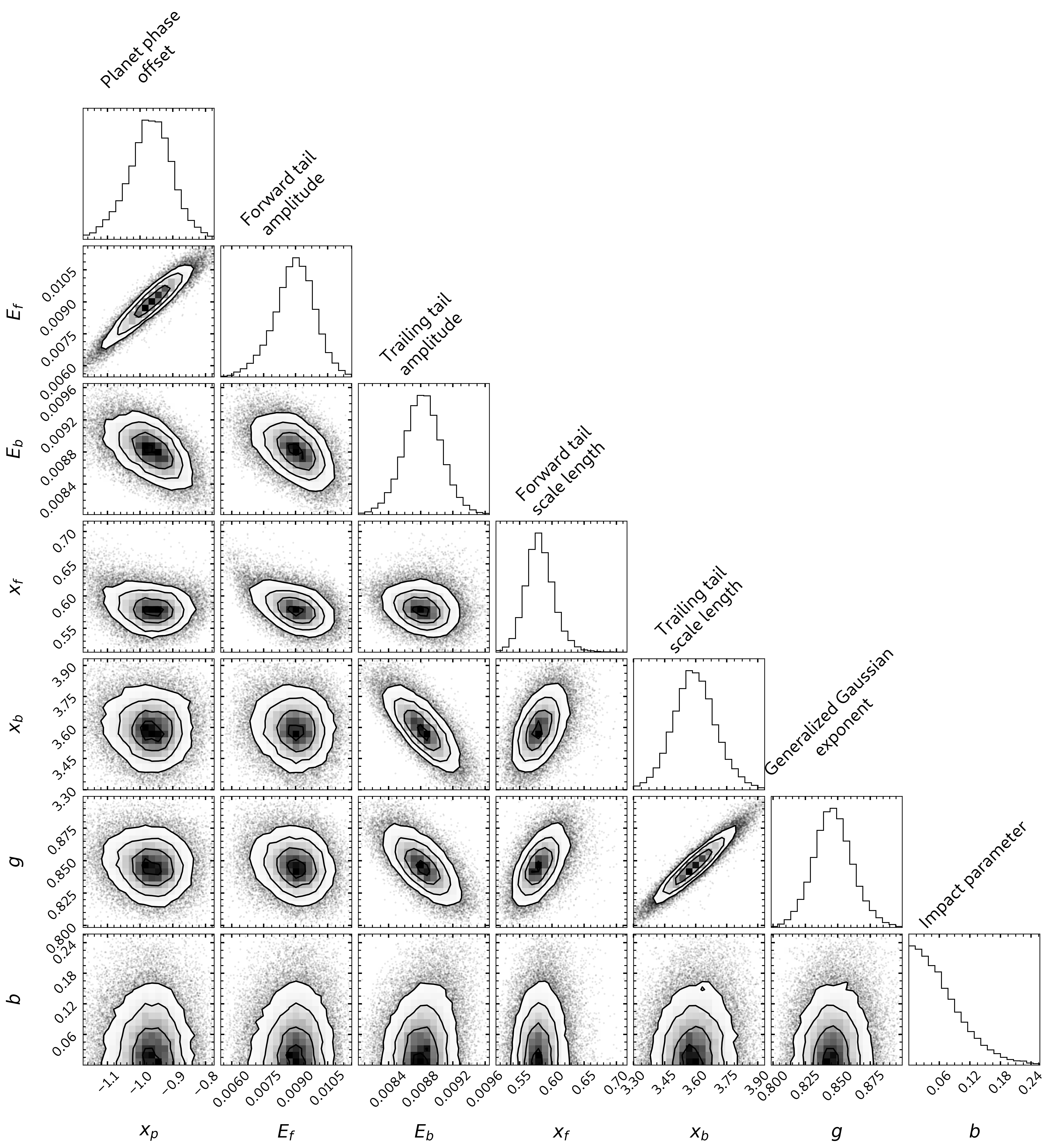}
    \caption{Corner plot of the posterior distribution to the fit to the mean transit profile of \system\,A's TESS light curve.}
    \label{fig:posterior}
\end{figure}

In this appendix, we present details of the fits to the mean transit profile observed by TESS. The corner plot displaying the joint posterior distribution is presented in Figure \ref{fig:posterior}, while the description and summary statistics of each fitted parameter are tabulated in Table \ref{table:mean_transit_fit}. 

We have previously fitted a model that allows for varying Generalized-Gaussian exponents in each tail (i.e., $g_f$ and $g_b$). The resulting best fit to the mean transit profile yielded $g_f = 0.89\pm0.11$ and $g_b = 0.84\pm0.02$. The values of these exponents are consistent with one another as well as the value of $g=0.85\pm0.01$ from the single-exponent model. Furthermore, adopting the two-exponent model resulted in an increase of Bayesian Information Criterion (BIC) by $\Delta$BIC $\simeq +7.5$, disfavoring the use of an additional exponential in our adopted model.

We additionally fitted a model which superimposes upon $\mathcal{Y}(d)$ the transit profile of a solid-body planet, with planetary radii sampled from uniform distribution of $R_p/R_{*}\sim\mathcal{U}(10^{-5}, 1)$. From the fits, we identified an upper limit of $R_p/R_{*}\simeq0.03$ or $R_{p}\simeq~3.6~R_{\oplus}$, above which the attenuation by the solid body resulted in progressively poorer fits to the observed mean transit profile. The inclusion of a solid body transit resulted in a statistically significant improvement in the fit, as indicated by a lower BIC value of $\Delta$BIC $\simeq -15.12$. However, the improvements in fits were focused at the transit minimum, which shows the strongest variations from transit to transit. Therefore, it is more likely that the model is overfitting the observed data, rather than a solid body transit signal being truly present.

\begin{table}[ht]
    \centering
    \caption{Best-fit parameters for the model of the mean transit profile (Equation \ref{eqn:mean_transit_fit}) when fit to the mean transit profile from TESS data. Quantities related to length and position are in units of $R_{\star}$. The prior distribution for each parameter is uniform (denoted as $\mathcal{U}$) across an interval whose boundaries are specified by the quantities in parentheses. Uncertainties of the median posterior values correspond to the 16$^{\mathrm{th}}$ and 84$^{\mathrm{th}}$ percentiles, respectively.}\label{table:mean_transit_fit}
    \begin{tabular}{cp{4.4cm}cp{3.2cm}}
        \toprule
         & \centering\arraybackslash Definition & Prior & \centering\arraybackslash Posterior Median  \\ 
        \midrule
        $x_p$ & \centering\arraybackslash Angular location of planet relative to transit minimum&$\mathcal{U}(-2.0, 1.2)$&\centering\arraybackslash$-0.97^{+0.07}_{-0.06}$ \\
        \hline
        $E_f$ & \centering\arraybackslash Forward tail amplitude &$\mathcal{U}(0, 0.1)$&\centering\arraybackslash$0.009\pm0.001$ \\
        \hline
        $E_b$ & \centering\arraybackslash Trailing tail amplitude &$\mathcal{U}(0, 0.1)$&\centering\arraybackslash$0.0088\pm0.0002$ \\
        \hline
        $x_f$ & \centering\arraybackslash Forward tail scale length &$\mathcal{U}(0.15, 5)$&\centering\arraybackslash$0.58\pm0.02$ \\
        \hline
        $x_b$ & \centering\arraybackslash Trailing tail scale length &$\mathcal{U}(0.15, 5)$&\centering\arraybackslash$3.6\pm0.1$ \\
        \hline
        $g$ & \centering\arraybackslash Generalized-Gaussian exponent &$\mathcal{U}(0.25, 5.25)$&\centering\arraybackslash$0.85\pm0.01$ \\
        \hline
        $b$ & \centering\arraybackslash Impact parameter &$\mathcal{U}(0,1)$&\centering\arraybackslash$0.05^{+0.02}_{-0.05}$ \\
        \bottomrule
    \end{tabular}
\end{table} 

\section{Fits to Individual Transits} \label{appendix:individual_transits}
\begin{figure}
    \centering
	\includegraphics[width=0.5\columnwidth]{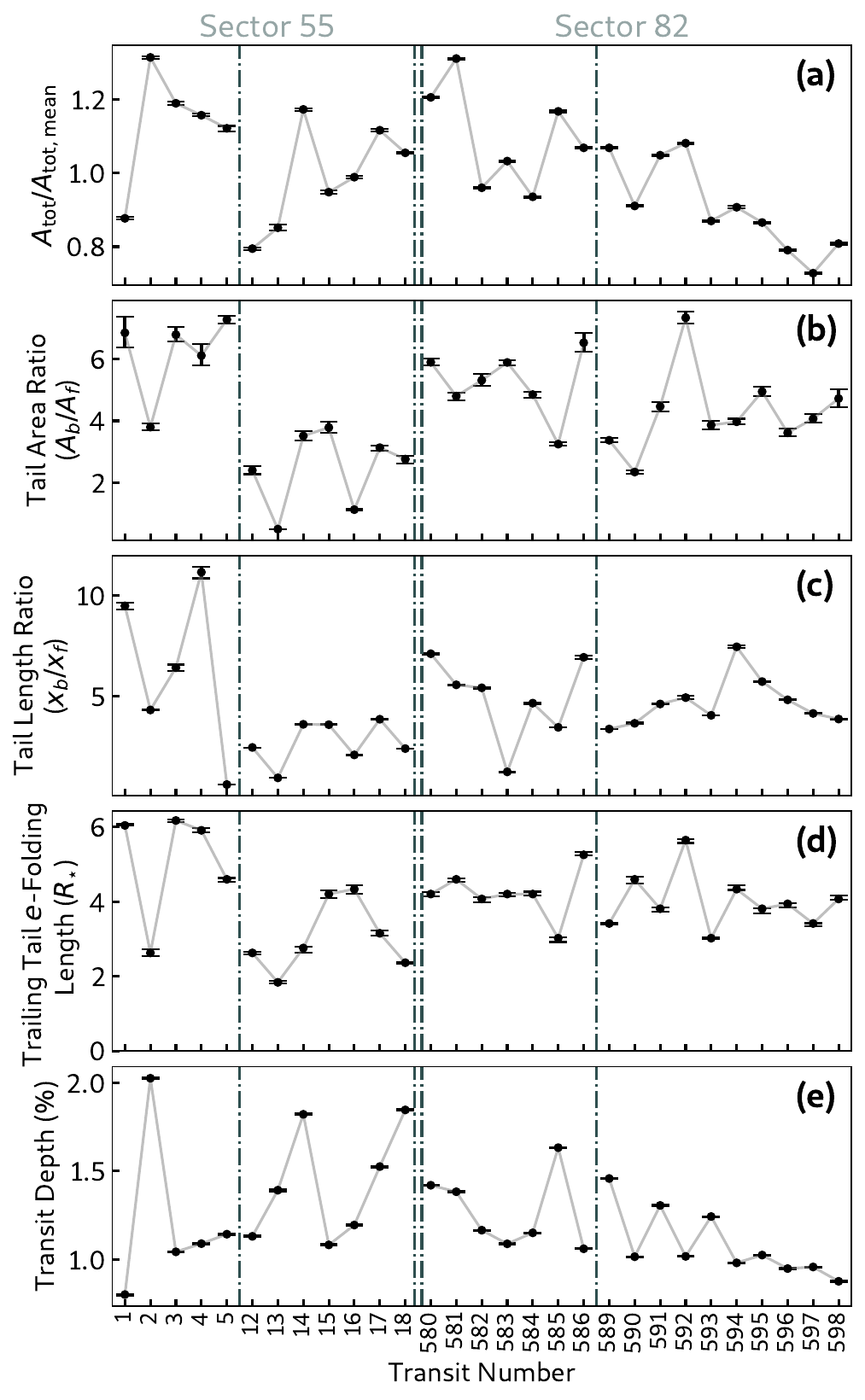}
    \caption{ Measured properties from the fitting of individual transits from the TESS light curves in Fig. \ref{fig:montage}, with a similar labeling scheme for the transit number. The vertical lines indicate gaps within the light curve; double lines indicate the gap between Sectors 55 and 82. The errors are computed from the 16$^{\mathrm{th}}$ and 84$^{\mathrm{th}}$ percentiles of the posterior distribution from model fits of $\mathcal{Y}(d)$. (a) The total integrated area under the extinction profiles of each transit ($A_{\mathrm{tot}}$), normalized by the corresponding value from the mean transit profile ($A_{\mathrm{tot, mean}}$). (b) Ratio of the integrated area under the trailing tail's extinction profile with that from the leading tail. (c) Ratio of Generalized Gaussian scale lengths between the trailing and leading tails. (d) The trailing tails' $e$-folding length in units of $\phi$, assuming an exponential decay. (e) Transit depth, measured as the minimum flux level in the transit profile.}
    \label{fig:transit_stats}
\end{figure}

In addition to fitting the mean transit profile of \system\,A's TESS light curve (Section \ref{section:light_curve}), we also fitted individual transits in the light curve using $\mathcal{Y}(d)$ (Eqn. \ref{eqn:mean_transit_fit}). The fits to the individual transits are presented as the red curves in Fig. \ref{fig:montage}. By fitting $\mathcal{Y}(d)$ to each transit, we measured the properties across the leading and trailing tails from transit to transit, such as the combined attenuation per transit ($A_{\mathrm{tot}}$, Fig. \ref{fig:transit_stats}a), the relative amount of attenuation between the trailing and leading tails ($A_{\mathrm{b}}/A_{\mathrm{f}}$, Fig. \ref{fig:transit_stats}b), and the ratio of the tail scale lengths between the trailing and leading tails ($x_{\mathrm{b}}/x_{\mathrm{f}}$, Fig. \ref{fig:transit_stats}c). For completeness, we calculated additional metrics per transit, including the angular distance required for the fitted attenuation profile for the trailing tail to reduce to $1/e$ (Fig. \ref{fig:transit_stats}d) and the depth of each transit (Fig. \ref{fig:transit_stats}e).

\section{Varying Simulation Initial Launch Parameters} \label{appendix:alternative_launches}
\begin{figure}
    \centering
	\includegraphics[width=1\columnwidth]{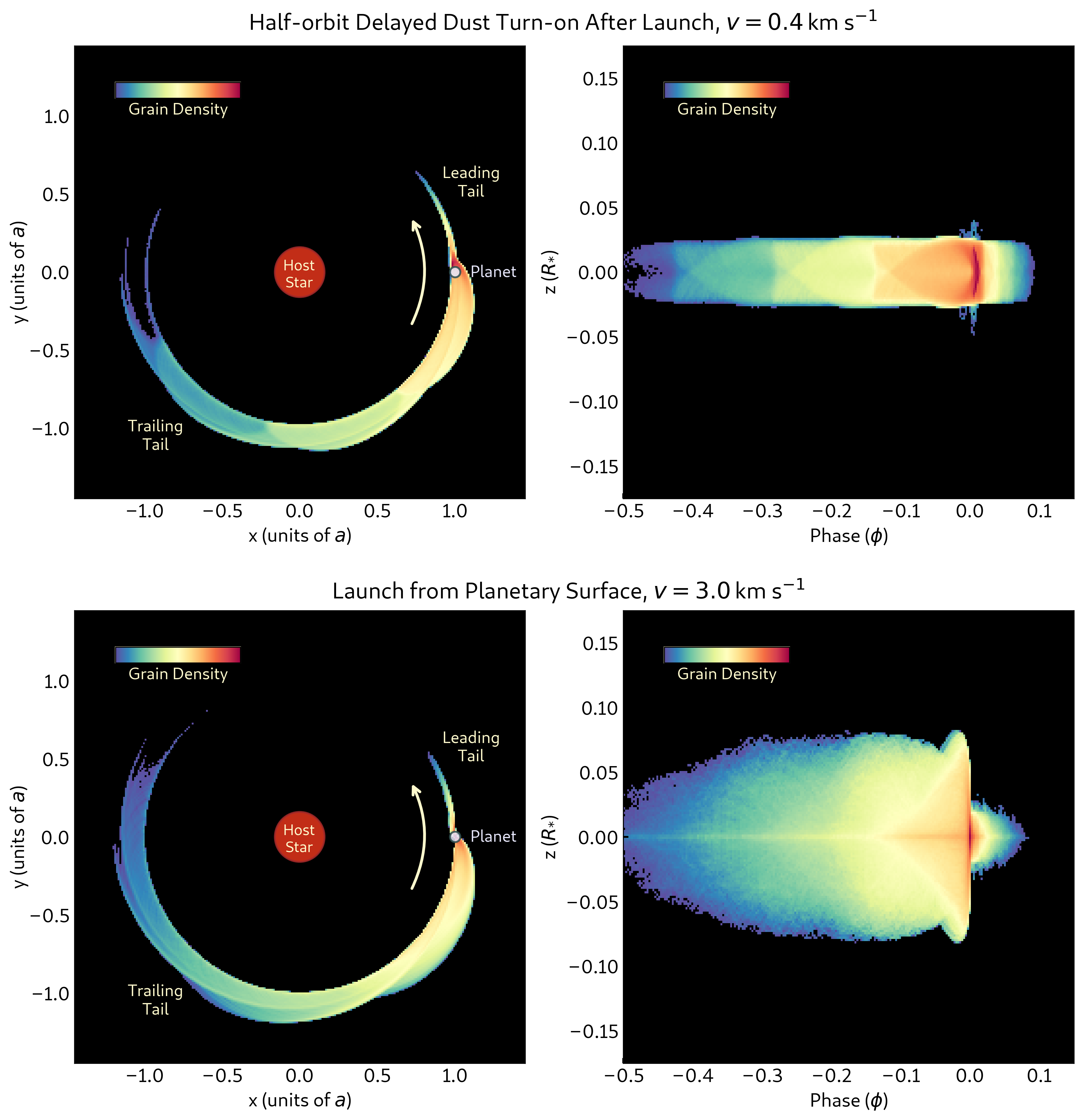}
    \caption{Same as Fig. \ref{fig:tail_sim}, but using different initial launch parameters for the dust grains. In the top row, the grains are simulated to form only at a distance past the Hill sphere by setting $\beta=0$ for half of the first orbit after release. In the bottom panel, the grains are released from the surface of a $0.02M_{\oplus}$ planet with a radius of $2000\,$km with a velocity corresponding to the planet's escape velocity (in isolation) of $3\,$km s$^{-1}$.}
    \label{fig:sim_alternative}
\end{figure}

In this appendix, we show results from repeating the simulations in Section \ref{section:numerical} but using different initial conditions for the release of the dust grains. In the top row of Fig. \ref{fig:sim_alternative}, we release dust grains at the Hill surface but set $\beta=0$ for all grains during the first half-orbit of the grains' trajectories (i.e., about 15 hrs). During this time window, a significant fraction of the grains may freely populate the leading tail. As the effects of $\beta$ are introduced after the first half-orbit, grains are blown into the trailing tail at distances sufficiently far from the planet, such that they do not experience the vertical launch assist described in Section \ref{section:numerical}. As a result, the trailing tail maximum height is now much lower at $\sim0.02\,R_{\star}$, implying that the dust tails will need to have moderately large optically depths ($\tau\sim2$) in order to reproduce the observed $1-2\%$ transit depths. This launch scenario would be incompatible if a deeper (e.g., $\sim3\%$) transit is ever observed for \system\,Ab.

In the bottom row of Fig. \ref{fig:sim_alternative}, we present the scenario whereby dust grains are launched from the planetary surface at a velocity equal to the planet's escape velocity of 3 km s$^{-1}$. While this scenario successfully reproduces the requisite tail heights, the assumption of such a high instantaneous launch speed at the surface is incommensurate with a Parker wind scenario that assumes gradual acceleration by gas pressure.

\begin{acknowledgments}
Funding for the TESS mission is provided by the NASA Explorer Program. We acknowledge the use of public TESS data from pipelines at the TESS Science Office. This paper includes data collected by the TESS mission that are publicly available from the Mikulski Archive for Space Telescopes (MAST).
M.H. acknowledges support from NASA grant 80NSSC24K0228. 
KAC and CNW acknowledge support from the TESS mission via subaward s3449 from MIT. 
The postdoctoral fellowship of KB is funded by F.R.S.-FNRS grant T.0109.20 and by the Francqui Foundation.
DRC acknowledges partial support from NASA Grant 18-2XRP18\_2-0007.
This work is partly supported by JSPS KAKENHI Grant Number JP24H00017
and JSPS Bilateral Program Number JPJSBP120249910.
This research was carried out in part at the Jet Propulsion Laboratory, California Institute of Technology, under a contract with the National Aeronautics and Space Administration (80NM0018D0004).
This work makes use of observations from the LCOGT network. Part of the LCOGT telescope time was granted by NOIRLab through the Mid-Scale Innovations Program (MSIP). MSIP is funded by NSF. 
This research has made use of the Exoplanet Follow-up Observation Program (ExoFOP; DOI: 10.26134/ExoFOP5) website, which is operated by the California Institute of Technology, under contract with the National Aeronautics and Space Administration under the Exoplanet Exploration Program. 
This paper is also based on observations made with the MuSCAT instruments, developed by the Astrobiology Center (ABC) in Japan, the University of Tokyo, and Las Cumbres Observatory (LCOGT). MuSCAT3 was developed with financial support by JSPS KAKENHI (JP18H05439) and JST PRESTO (JPMJPR1775), and is located at the Faulkes Telescope North on Maui, HI (USA), operated by LCOGT. MuSCAT4 was developed with financial support provided by the Heising-Simons Foundation (grant 2022-3611), JST grant number JPMJCR1761, and the ABC in Japan, and is located at the Faulkes Telescope South at Siding Spring Observatory (Australia), operated by LCOGT.
Some of the data presented herein were obtained at Keck Observatory, which is a private 501(c)3 non-profit organization operated as a scientific partnership among the California Institute of Technology, the University of California, and the National Aeronautics and Space Administration. The Observatory was made possible by the generous financial support of the W. M. Keck Foundation. The authors wish to recognize and acknowledge the very significant cultural role and reverence that the summit of Maunakea has always had within the Native Hawaiian community. We are most fortunate to have the opportunity to conduct observations from this mountain.  
\end{acknowledgments}

\facilities{TESS, LCOGT}

\software{AstroImageJ \citep{Collins:2017}, TAPIR \citep{Jensen:2013}, Dynesty \citep{Speagle_2020}}

\bibliography{sample631}{}
\bibliographystyle{aasjournal}

\end{document}